\documentclass[epsfig,12pt]{article}
\usepackage{epsfig}
\usepackage{graphicx}
\usepackage{rotating}

\usepackage{latexsym}
\usepackage{amsmath}
\usepackage{amssymb}
\usepackage{relsize}
\usepackage{geometry}
\usepackage{color}
\usepackage{bm}

\def\beq{\begin{equation}}
\def\eeq{\end{equation}}
\def\beqn{\begin{eqnarray}}
\def\eeqn{\end{eqnarray}}

\newcommand{\ntwo}{${\mathcal N}=2\;$}

\newcommand{\ntt}{${\mathcal N}=(2,2)\,$}

\newcommand{\cell}{{\mathcal L}}

\newcommand{\pt}{\partial}

\newcommand{\gsim}{\lower.7ex\hbox{$
\;\stackrel{\textstyle>}{\sim}\;$}}
\newcommand{\lsim}{\lower.7ex\hbox{$
\;\stackrel{\textstyle<}{\sim}\;$}}

\def\beqn{\begin{eqnarray}}
\def\eeqn{\end{eqnarray}}

\def\beq{\begin{equation}}
\def\eeq{\end{equation}}
\def\ba{\beq\new\begin{array}{c}}
\def\ea{\end{array}\eeq}

\newcommand{\ntwot}{${\mathcal N}= \left(2,2\right) $ }

\newcommand{\mc}[1]{\mathcal{#1}}

\newcommand{\lgr}{\left\lgroup}
\newcommand{\rgr}{\right\rgroup}

\def\slashed#1{\setbox0=\hbox{$#1$}             
   \dimen0=\wd0                                 
   \setbox1=\hbox{/} \dimen1=\wd1               
   \ifdim\dimen0>\dimen1                        
      \rlap{\hbox to \dimen0{\hfil/\hfil}}      
      #1                                        
   \else                                        
      \rlap{\hbox to \dimen1{\hfil$#1$\hfil}}   
      /                                         
   \fi}                                        %

\newcommand{\mbps}{m_{\text{\tiny BPS}}}
\newcommand{\W}{\mathcal{W}}
\newcommand{\hsigma}{{\hat{\sigma}}}

\begin{document}

\hyphenation{con-fi-ning}
\hyphenation{Cou-lomb}
\hyphenation{Yan-ki-e-lo-wicz}
\hyphenation{di-men-si-on-al}


\begin{titlepage}

\begin{flushright}
FTPI-MINN-11/09, UMN-TH-2942/11\\
April 24/2011
\end{flushright}

\vspace{1.1cm}

\begin{center}
{  \Large \bf  BPS Spectrum of Supersymmetric CP($N-1$) \\[3mm]
		Theory with \boldmath{$\mc{Z}_N$} Twisted Masses}
\end{center}
\vspace{0.6cm}

\begin{center}

 {\large
 \bf   Pavel A.~Bolokhov$^{\,a,b}$,  Mikhail Shifman$^{\,a}$ and \bf Alexei Yung$^{\,\,a,c}$}
\end {center}

\begin{center}

$^a${\it  William I. Fine Theoretical Physics Institute, University of Minnesota,
Minneapolis, MN 55455, USA}\\
$^b${\it Theoretical Physics Department, St.Petersburg State University, Ulyanovskaya~1, 
	 Peterhof, St.Petersburg, 198504, Russia}\\
$^{c}${\it Petersburg Nuclear Physics Institute, Gatchina, St. Petersburg
188300, Russia
}
\end{center}

\vspace{0.7cm}

\begin{center}
{\large\bf Abstract}
\end{center}

\hspace{0.3cm}
	We revisit  the BPS spectrum of 
	the supersymmetric CP($N-1$) two-dimensional model with $\mc{Z}_N$-symmetric twisted masses $m_l$
	($l=0,1, ..., N-1$).
	A related issue we address is that of the curves of marginal stability (CMS) in this theory. 
	Previous analyses were incomplete.
	We close the gap by exploiting a number of consistency conditions. In particular, we amend the Dorey formula
	for the BPS spectrum.
	Our analysis is based on the exact Veneziano--Yankielowicz-type superpotential and on
	the strong-coupling spectrum of the theory found from the mirror representation at small masses, $|m_l|\ll\Lambda$.
	We show that at weak coupling the spectrum, with necessity, must include $ N - 1 $ BPS towers of states, instead
	of just one, as was thought before. 
	Only one of the towers is seen in the quasiclassical limit.
	We find the corresponding CMS for these towers, and argue that in the large-$ N $ limit
	they become circles, filling out a band on the plane of a single mass parameter of the model at hand. 
	Inside the CMS, $N-1$ towers collapse into $N$ stable states.
\vspace{2cm}

\end{titlepage}

\newpage

\tableofcontents

\newpage

\section{Introduction}
\setcounter{equation}{0}

	The two-dimensional  \ntwot CP($N-1$) model with twisted masses exhibits  
	a remarkable similarity to certain four-dimensional gauge theories. The underlying reason was revealed in \cite{SYmon,HT2}
	where the supersymmetric CP($N-1$) model was demonstrated to emerge as a low-energy world-sheet theory
	on the non-Abelian strings, see \cite{HT1,ABEKY,SYmon,HT2}. The coincidence between the BPS spectrum
	of the CP($N-1$) model and that of \ntwo SQCD in a quark vacuum observed in \cite{Dor} received a natural explanation.
	Namely, these two spectra, in fact, describe the same states but from two different 
	perspectives: viewed from the bulk and from the world sheet. 
	
	Superalgebra in the CP($N-1$) model is centrally extended, with the central charge containing a topological term and
	$(N-1)$ Noether charges. The former includes a canonical contribution and an anomalous one, 
	which is especially obvious in the limit of the vanishing twisted masses   (e.g. \cite{ls1}). The contributions of the 
	 Noether charges are proportional to the twisted masses. 
	 
	 The masses of the BPS-saturated states reduce to the corresponding central charges. However, not every 
	  state is realized in the theory. The BPS spectrum depends on the value of the twisted masses
	 $m_l$ ($l=0,1, ..., N-1$). If $ |m_l| \:\gg\: \Lambda$, we are at weak coupling. In this regime the BPS spectrum
	 contains states characterized by the topological charge $T \:=\: \pm1$, with infinite towers of heavier states with nonvanishing 
	 U(1) charges built on them. Not all states possible algebraically are dynamically realized as stable states in the spectrum.
	 If $|m_l| \:\ll\: \Lambda$, we are at strong coupling. The BPS spectrum of elementary kinks
         shrinks down to $N$ stable states.
	 
	 A general and detailed discussion of the BPS mass spectrum in the  CP($N-1$) model with the twisted masses
	 was undertaken in \cite{HaHo,Dor} on the basis of the 
	 Veneziano--Yankielowicz-type\,\footnote{The Veneziano--Yankielowicz superpotential in the CP$(N-1)$ models
	 with no twisted masses was originally derived, in terms of twisted superfields,
	 in \cite{AdDVecSal,ChVa,W93}.} 
	 superpotential  
	 \cite{VYan} augmented by the 
	 twisted mass terms \cite{HaHo}, in conjunction with a quasiclassical analysis.
	  A concrete implementation of the  spectrum of  the dynamically realized  
	 BPS states on both sides of the CMS (as well as the CMS itself)
	 in the CP(1) model  was constructed in  \cite{ls1}. 
	 For $N \,>\, 2$ in the case of $\mc{Z}_N$-symmetric twisted masses
	 the issue of the BPS spectrum and the CMS separating the strong and weak-coupling domains was addressed
	 in \cite{Olmez} and, later, in   \cite{Bolokhov:2010hv}. The latter two studies were based on the Dorey formula
	 which, as we will show in this paper, is by far incomplete. The advantage of the $\mc{Z}_N$-symmetric
	 twisted masses compared to a generic set is rather obvious. Generally speaking, we have $N-1$ twisted masses
	 implying the parameter space of the complex dimension $N-1$ (in this case it is more appropriate to speak
	 of the decay walls rather than curves). Starting from $N \,=\, 3$ and higher, with generic masses,
	 explicit  analytic determination of the BPS spectra and CMS, even if possible, 
	   ceases to be instructive. 	On the other hand, with the $\mc{Z}_N$-symmetric twisted masses,
	   the model depends on a single complex mass parameter, everything becomes simple, and the would-be decay walls
	   reduce to a set of CMS in the complex plane of $m_0$, a single complex parameter 
	   defining all twisted masses. 
	On the physical side the advantage of the $\mc{Z}_N$-symmetric
	   choice is also obvious: with this choice the twisted masses preserve the $\mc{Z}_N$ subgroup of the 
	   $ \mc{Z}_{2N} $ symmetry of the massless model, which is a remnant of the axial $R$ symmetry. 
	Physics of the transition from weak to strong coupling becomes transparent.
	   
	   Our goal here is to work out a complete solution of this question. Technically the problem is due to the fact that 
	   the Veneziano--Yankielowicz (VY) superpotential is a multivalued function and, in calculating the BPS spectrum, 
	   one needs to carefully 
	   analyze its branches in order to select those
	   which can be dynamically realized.
	    We add two crucial ingredients missing in \cite{Dor}. First, in the small-$|m_l|$ limit the 
	    BPS spectrum is known \cite{Shifman:2010id} from the Hori--Vafa representation \cite{MR1}.
	In other words, the mirror symmetry is instrumental  in finding the 
	BPS spectrum.
		At strong coupling  only $ N-1 $ stable states exist.
	These are the states that become massless at the Argyres--Douglas points \cite{AD},
	at which vacua collide. In our case, the $\mc{Z}_N$-symmetric masses, all vacua collide simultaneously.
	The vanishing mass requirement at this point, in conjunction with small mass formula,
	enables us to resolve the ambiguity in the VY superpotential and determine
	the spectrum of masses of stable BPS states both outside and inside the CMS in its entirety.
	We arrive at the conclusion  that the spectrum found in \cite{Dor} and consisting
	of a single BPS tower (for kinks interpolating between a given pair of vacua)
	is not capable of satisfying all the above requirements.
	We argue that the theory must instead have $ N - 1 $ towers, each of which is described
	by its own U(1) quantum number.
	We find this natural in view of the fact that the global SU($N$) symmetry is broken
	down to U(1)$^{N-1}$ by the twisted masses.
	For each of these U(1)'s there is a tower of states arising from quantization.
	Only one tower is distinctly seen in the quasiclassical treatment, however, and that makes it special.
	That is the tower described in \cite{Dor}. All others are not resolved:
	in the quasiclassical limit which corresponds to large $ | m_l | $ and large U(1) quantum numbers
	all $ N-1 $ towers fuse with each other.
	
	With the BPS spectrum in hands we pass to the second stage: 
	determination of  the curves of marginal stability  in the same CP($N-1$) model
	with $\mc{Z}_N$-symmetric twisted masses.
	For each of the BPS towers there must be a curve on which the relevant states decay.
	Altogether, we deal here with $N-1$ curves of marginal stability.
	We find that the curve corresponding to the special ``quasiclassical'' tower 
	always passes through the Argyres--Douglas point. 
	For this reason we will call it {\em primary}.
	The primary curve is the innermost curve on the complex plane of 
	$m_0$, inside which only strong-coupling states are stable.
	Other, secondary, curves are larger in size and are (typically) near-perfect circles. 	
	When passing from the weak-coupling domain into the strong-coupling one, the 
	towers of states decay on these curves,  one by one.
	Finally, 
	we consider the large-$ N $ limit of the theory and show that in this limit 
	all CMS tend to round circles,
	with radii in the interval 
\beq
	 1 ~~\leq~~ |m|\,/\,\Lambda ~~\leq~~ e^2\,, \qquad\qquad (\, e ~=~ 2.71828\,...\,)
\eeq
	  in units of the 
	low-energy scale $ \Lambda $.
	In the limit of very large $ N $ the curves of marginal stability  completely fill  this interval, forming
	a round band.
	
	Organization of the paper is as follows. In Sec. \ref{super} we review the gauged formulation of the 
	model and present the VY superpotential with the twisted mass included. In Sec. \ref{mitr}
	we consider predictions for the BPS spectrum at small $|m_0|$ following from the mirror symmetry.
	Section \ref{adpo} is devoted to the Argyres--Douglas (AD) point. In Sec. \ref{sbps} we obtain the complete BPS
	spectrum both at strong and weak coupling based on the VY superpotential and the conditions outlined above.
	Construction of $N-1$ curves of marginal stability is the subject of Sec. \ref{curves}. Section \ref{conclu} presents a
	summary of our results.

\section{Exact Superpotential}
\label{super}
\setcounter{equation}{0}

In the gauged formulation, the bosonic part of the Lagrangian of the \ntt super\-symmetric CP($N-1$) theory can be obtained
from four-dimensional SQED with an $N$-plet chiral superfield, by a dimensional reduction to two dimensions
\cite{W93,HaHo},
\begin{align}
\cell & ~~=~~ 
	\frac{1}{e_0^2} \lgr \frac{1}{4} F_{\mu\nu}^2 ~+~ \left|\pt_\mu\sigma\right|^2 ~+~ \frac{1}{2}D^2 \rgr
 	~~+~~ \left|\nabla_\mu n^i\right|^2 
	\nonumber
	\\[3mm]
&
	~~+~~ i\,D\left( |n_i|^2 \;-\; 2\beta \right)
	~~+~~ 2\,\sum_i \Bigl| \sigma-\frac{m_i}{\sqrt 2} \Bigr|^2\, |n^i|^2\,.
\label{fullcpn}
\end{align}
in the strong-coupling limit $ e_0 \,\to\, \infty$. Here $e_0$ is the gauge coupling while
	$ m_i $ are the complex twisted mass parameters which can be introduced \cite{twistedm} via a
	background vector field in four dimensions. Moreover, $2\beta$ can be viewed as a coupling of the
	corresponding sigma model (sometimes $r \:\equiv\: 2\beta$ is used).

	We should mention that physically the mass parameters are not given by the masses $ m_l $ themselves,
	but, rather, by their differences $ m_l - m_k $ (or by $ m_l - m $, where $ m $ is the average mass).
	This is clear from Eq.~\eqref{fullcpn} as one can shift all masses by any value via a redefinition of
	$ \sigma $. 
	Thus, one can always impose the condition
\beq
	\sum_{l\,=\,0}^{N-1}\, m_l ~~=~~ 0\,.
	\label{mcondi}
\eeq

	With arbitrary choice of masses we break the global SU($N$) invariance of the model down to U(1)$^{N-1}$.
	We are interested in a special case when the masses are assumed to preserve a discrete $ \mc{Z}_{N} $ 
	subgroup 
	of the anomalous U(1)$_\text{R}$ symmetry.  To this end they must sit
	equidistantly on a circle,
\beq
\label{mcirc}
	m_l ~~=~~ m_0 \cdot e^{2 \pi i l / N}\,,\qquad l=0,1, ..., N-1\,,
\eeq
	with a single complex parameter $ m_0 $ defining all masses. The condition
	(\ref{mcondi}) is automatically met.
	We will see that in the CP$(N-1)$ model, the physical dependence of the theory is, in fact, on $ m_0^N $.

	The theory \eqref{fullcpn} classically has $ N $ vacua, which can be seen as solutions with all $ n_i $ but one 
	set to zero,
\begin{align}
	n_i & ~~=~~ (\, 0,~ ...,~ 1, ...,~ 0\, )\,,  	\qquad\qquad\qquad  k ~=~ 0,..., N-1\,,
\notag
	\\
	\sigma & ~~=~~ m_k \,.
\label{nvacua}
\end{align}
	Note that we choose to number both the masses and the vacua from $0$ to $N-1$.

	The chiral sector of this theory is described by  an exact superpotential of the
Veneziano--Yankielowicz type 
	\cite{VYan}.
	For a theory with twisted masses the 
Veneziano--Yankielowicz superpotential was 
	derived in \cite{AdDVecSal,ChVa,W93,HaHo}, 
	and is obtained by integrating out the $ n^l $ fields in Eq.~\eqref{fullcpn},
\beq
\label{Wfull}
	\W_\text{eff}(\hsigma) ~~=~~
		-\, i\, \tau \hsigma ~~+~~
		\frac{1}{2\pi} \sum_j (\hsigma - m_j)\, 
				      \left\{ \ln {\frac{\hsigma - m_j}{\mu}} ~-~ 1 \right\}\,.
\eeq
	The hat over $ \hsigma $ indicates that it is actually a (twisted) superfield.
	In passing from (\ref{fullcpn}) to (\ref{Wfull}) we rescaled $ \sigma $,
\beq
	\sigma ~~\to~~ \frac{\sigma}{\sqrt 2}\,.
	\label{resca}
\eeq
	In Eq.~\eqref{Wfull} $ \tau $ is the complexified coupling,
\beq
	\tau ~~=~~ i r ~+~ \frac{\theta}{2\pi}\,, \qquad\qquad\qquad   \text{with~~~} r ~~\equiv~~ 2\beta\,.
\eeq
	The ultraviolet cut-off scale $ \mu $ can be traded for the dynamical scale $ \Lambda $,\footnote{Without a loss
	of generality one can always assume $\Lambda$ to be real and positive. We will stick to this convention. 
	A possible phase of $\Lambda$ is absorbed in $m_0$.}
\beq
	\mu ~~=~~ \Lambda\, e^{2\pi r/N}\,,
\eeq
where $r$ in the exponent on the right-hand side corresponds to the same normalization point $\mu$.
Then, the first term on the right-hand side of (\ref{Wfull}) disappears while $\mu$ in the argument of logarithms is replaced by $\Lambda$.
The CP$(N-1)$ model is asymptotically free.

	Now, Witten's formula \cite{W93} for the positions of vacua is
	\beq
	\prod_{l\,=\,0}^{N-1} \, \left( \sigma \,-\, m_l \right) ~~=~~ \Lambda^N\,.
	\label{Wf}
	\eeq
	This is the equation for critical points of $\W_\text{eff}(\sigma)$. With the set of masses (\ref{mcirc})
	Eq. (\ref{Wf}) implies that the
	vacua of this theory are at
\beq
\label{sigvac}
	\sigma_p ~~=~~ \sqrt[N] { \Lambda^N \,+\, m_0^N } \cdot e^{ 2\pi i p / N }\,, \qquad p ~=~ 0,\,...,\, N-1\,.
\eeq
	Altogether, we have $N$ vacua, in full accord with Witten's index.
	
	If $ | m_ 0 | $ is taken to be large, then it dominates over $ \Lambda $ in \eqref{sigvac},
	and the vacua take their classical values \eqref{nvacua},
\beq
	\sigma_p ~~\approx~~ m_p\,, \qquad\qquad\quad p ~=~ 0,\,...,\,N-1\,.
\eeq
	In the future, for determination of the BPS spectrum, we will need the values of the superpotential
	in the vacuum. It is not difficult to get
\beq
\label{Wvac}
	\W_\text{eff} ( \sigma_p ) ~~=~~ 
		-\, \frac{1}{2\pi}\,  
                \Bigl\{\, N\, \sigma_p ~+~ \sum_j\, m_j\, \ln \,\frac{\sigma_p - m_j}{\Lambda} \,\Bigr\}\,.
\eeq
Various choices of the branches of the logarithms above correspond to distinct values of the Noether U(1) charges.

In deriving expression (\ref{Wvac}) we used the Witten's relation (\ref{Wf}). With our normalization, the
	general formula for the mass of an {\em elementary} BPS state reduces to the difference of 
	superpotentials in two neighboring vacua, 
\beq
\label{mbpsgen}
	\mbps ~~=~~ \big|\, \W_\text{eff} ( \sigma_{p+1} ) ~~-~~ \W_\text{eff} ( \sigma_p ) \,\big|\,.
\eeq
Note that the elementary kinks are obtained if one interpolates between the neighboring vacua. 
The result is the same independently of 
which pair of neighbors we pick up. This is due to the  $\mc{Z}_N$ symmetry of the model.

	With the $ \mc{Z}_N $-symmetric  masses \eqref{mcirc}, the theory at quantum level
	retains the $ \mc{Z}_{N} $ symmetry, which the masses do not break.
	This symmetry manifests itself in the invariance of the spectrum 
	to the choice of vacua in \eqref{mbpsgen}.
	From now on we will  choose the vacua $ \sigma_0 $ and $ \sigma_1 $ as representatives
	and focus on the masses of kinks interpolating between the two,
\beq
\label{mbpsmain}
	\mbps ~~=~~ \big|\, \W_\text{eff} ( \sigma_1 ) ~~-~~ \W_\text{eff} ( \sigma_0 ) \,\big| 
\eeq
	(in bulky expressions, we will  omit the absolute value sign, keeping it in mind; 
	certainly, $ \mbps $ is always a positive quantity).

	In the quasiclassical limit $ |\Delta m| \,\gg\, \Lambda $ the leading contribution to the mass
	is given by the dominant logarithm in expression \eqref{mbpsmain},
\beq
\label{wkink}
	\W_\text{eff} ( \sigma_1 ) ~~-~~ \W_\text{eff} ( \sigma_0 ) ~~\sim~~
		\frac{N}{2\pi}\,
		\Delta m \cdot \ln \frac{| m_0 |}{\Lambda} ~~=~~  r \cdot \Delta m\,, 
\eeq
	where $ \Delta m ~=~ m_1 \,-\, m_0 $.

	It is well-known that the 
Veneziano--Yankielowicz potential, being a multibranch function,
	is too ambiguous.
	The degree of ambiguity of expression \eqref{mbpsmain} is determined by the logarithms in \eqref{Wvac},
	and can be symbolized by a linear combination of the masses $ m_j $ with arbitrary integer coefficients
\beq
\label{amb}
	\langle\text{integer}\rangle_j \cdot m_j \,.
\eeq
	If all these multiplicities were physical, one would have a set of $ \mc{Z}^N $ states in the spectrum, which
	is certainly not what is expected.
	Selection rules need to be formulated in order to restrict the set of the BPS states
	that actually exist.
	We postpone the formulation of these rules until further, while for now do a simple mathematical
	trick which reduces the amount of ambiguity present in Eq.~\eqref{mbpsmain}.
	Our goal is to turn \eqref{mbpsmain} being a function of all masses and two vacua into a function
	of a single parameter $ m_0 $.

	Let us pull out a factor of $ e^{2\pi i / N} $ from each term in 
	$ \W_\text{eff} ( \sigma_1 ) $ which originally looks as,
\beq
\label{Wsigone}
	\W_\text{eff} ( \sigma_1 ) ~~=~~ 
		-\, \frac{1}{2\pi}\,  
                \Bigl\{\, N\, \sigma_1 ~+~ \sum_j\, m_j\, \ln \,\frac{\sigma_1 - m_j}{\Lambda} \,\Bigr\}\,.
\eeq
	Both terms in Eq.~\eqref{Wsigone} do contain this factor.
	This move turns $ \sigma_1 $ into $ \sigma_0 $, while shifts the numeration of masses in the sum.
	To keep the numeration of masses in the sum consonant with the logarithms, we also pull out
	$ e^{2\pi i / N} $ from the argument of the logarithm.
	This constant addition vanishes when summed with $ \sum\, m_j $,
	while $ \sigma_1 $ inside the logarithm again turns into $ \sigma_0 $.
	Effectively one arrives at 
\beq
	\W_\text{eff}(\sigma_1) ~~\stackrel{?}{=}~~ e^{2\pi i / N}\, \W_\text{eff}(\sigma_0)\,.
	\label{quesm}
\eeq

The question mark in (\ref{quesm}) reminds us that
	this is not quite the whole story. We were not very careful with the phases
	of the logarithms, and could have easily missed (and actually did) some $ 2 \pi i $.
	This owes to the fact that $ \ln\,a b ~=~ \ln a ~+~ \ln b $ only modulo $ 2\pi i $.
	But such an omission can just as well be merged into the general ambiguity \eqref{amb}
	of the expression \eqref{mbpsmain}.
	We can fuse this ambiguity into a single linear combination $ i\, \vec{N} \cdot \vec{m} $, where $ \vec{N} $
	is an arbitrary constant integer vector, 
$\vec{N} ~=~ (n_0,~ ...,~ n_{N-1})$ 
	with integer $n_i$ and
$\vec{m} ~=~ (m_0,~ ...,~ m_{N-1})$. 
	Now we rewrite \eqref{mbpsmain} as,
\beq
\label{mspec}
	\mbps ~~=~~ U_0 (m_0) ~~+~~ i\, \vec{N} \cdot \vec{m}\,,
\eeq
	with an explicit function
\begin{align}
\label{unod}
	U_0 (m_0) & ~~=~~ 
	\\
\notag
	&\!\!\!\! -\, \frac{1}{2\pi} \lgr e^{2\pi i / N} \,-\, 1 \rgr 
	\biggl\{\, N \sqrt[N] { m_0^N \,+\, \Lambda^N }  ~+~
		\sum_j\, m_j\, \ln \, \frac{ \sqrt[N] { m_0^N \,+\, \Lambda^N } \,-\, m_j } { \Lambda} \,\biggr\}\,.
\end{align}
The branch of $U_0$ is fixed as follows: if $x$ is a real positive number, (i) $\sqrt[N] {x}$ is a real positive number; (ii)
$\ln x$ is a real number. Here $ m_j $ is meant to be a function of $ m_0 $ as well, via \eqref{mcirc}. 

	The important statements are,
\begin{itemize}
\item
	In Eq.~\eqref{mspec} all ambiguities related to the logarithms are referred to $ \vec{N} $;
\item
	Now $ U_0(m_0) $ is a fixed single-valued function of the complex parameter $ m_0 $
	in a certain region of the complex plane.

\item
	Vector $ \vec{N} $ in Eq.~\eqref{mspec} has a direct relation to the spectrum.
\end{itemize}
	Let us briefly comment on what is meant.
	The BPS spectrum exists everywhere on the complex plane of $ m_0 $, and both expressions \eqref{mbpsmain}
	and \eqref{mspec} must in principle describe it.
	But for that, they need to be made unambiguous at least in some region of the complex plane of $m_0$. 
	We claim that function $ U_0(m_0) $ is single-valued in a domain of the complex plane
	wide enough to unambiguously determine the spectrum. The latter will be described
	in terms of the vector $ \vec{N} $.
	Exactly how wide the domain of the parameter $ m_0 $ needs to be will be discussed in Section~\ref{sbps}.
	It will also become clear why Eq.~\eqref{mspec} is more directly related to the determination 
	of the spectrum than Eq.~\eqref{mbpsmain}.
	
	In this way we will arrive at our master formula.
	We will be able to determine such an expression for the spectrum which will be consistent 
	both at large and small $ |m| $.
	In particular, we will find that the prescription obtained in \cite{Dor}, 
\beq
\label{qclass}
	\vec{N} ~~=~~ (\, -\,n,~~~ n,~~~ 0,~~~ ...,~~ 0\, )\,, 
	\qquad\qquad\quad     n~\in~\mc{Z}\,,
\eeq
	is valid only approximately, in the quasiclassical limit, which corresponds 
	to large $ |m| $ {\it and} large excitation number $ n $.
	The result \eqref{qclass} was derived by quasiclassical quantization of the time rotation of 
  kinks in the U(1) factors of the global group, 
	and, therefore, must still be valid as an {\it asymptotics}.
	Below we argue, however, is that the description \eqref{qclass} is incomplete.

\section{Mirror Treatment}
\label{mitr}
\setcounter{equation}{0}

	Now we gradually pass to the discussion of what is known 
	about the strong-coupling regime.
	There are $ N $ BPS kinks.
	They can be seen in the mirror representation \cite{MR1,MR2}, 
\beq
\label{mirror}
	\W_\text{mirror}^\text{CP($N-1$)} ~~=~~
		-\, \frac{\Lambda}{2\pi}\, 
		\Bigl\{\, \sum_j X_j ~~+~~ \sum_j \frac{m_j}{\Lambda}\, \ln X_j \,\Bigr\}\,,
\eeq
	with
\beq
	\prod_j\, X_j ~~=~~ 1\,.
\eeq
	As was shown in \cite{Shifman:2010id}, one can determine the masses of all $ N $ kinks
	near the origin, $ | m_j | ~\ll~ \Lambda $,
\beq
\label{mirrorm}
	\mbps ~~\approx~~ \Big|\, \frac{N}{2\pi} \lgr e^{2\pi i / N} \,-\, 1 \rgr \Lambda
			   ~~-~~ i\, ( m_j \,-\, m ) \,\Big|\,,
\eeq
	where $ m $ is the average mass, vanishing in the $ \mc{Z}_N $ case.
	So, to the linear order in the mass parameter, one has $ N $ kinks with the masses given
	by a large $ \Lambda $ term, and the splittings determined by $ m_j $ themselves,
\beq
\label{smirror}
	\mbps ~~\approx~~ \Big|\, \frac{N}{2\pi} \lgr e^{2\pi i / N} \,-\, 1 \rgr \Lambda
			   ~~-~~ i\, m_j \,\Big|\,,
	\qquad\quad j~=~ 0,\,...,\, N-1\,.
\eeq

\section{Argyres--Douglas Point}
\label{adpo}
\setcounter{equation}{0}

	If all masses $ m_j $ sit on a circle the corresponding
	vacua will also be forced to sit on a circle, see
	\eqref{sigvac}.
	Therefore, only simultaneous collisions of all $N$ vacua $ \sigma_p $
	take place in the theory at hand.
	The 
Argyres--Douglas points \cite{AD}  correspond to 
\beq
	\sigma_p ~~=~~ 0\,.
\eeq
	This occurs whenever $ m_0^N ~=~ - \Lambda^N$, or, in the $ m_0 $ plane,
\beq
	m_0^\text{AD} ~~=~~ \Lambda \, e^{i \pi / N} \cdot e^{2\pi i l / N}\,,
	\qquad\qquad\quad 
	l ~=~ 0,\,...,\, N-1 \,.
\eeq
	Of these, the most convenient for us will be the two AD points closest to the
	real positive axis,
\beq
	m_0^\text{AD(I)} ~~=~~ \Lambda \, e^{i \pi / N}
	\qquad\quad
	\text{and}
	\qquad\quad
	m_0^\text{AD(II)} ~~=~~ \Lambda \, e^{- i \pi / N}\,.
\eeq

	The crucial observation about the AD point is that one of the $N$ soliton states
\eqref{smirror} becomes massless
	at that location in the complex $m_0$ plane.
	Briefly, if all the vacua merge at the AD point, then Eq.~\eqref{mbpsmain} tells one 
	that one of the kinks becomes massless 
\beq
	\mbps ~~=~~ \W_\text{eff}(\sigma_1) ~-~ \W_\text{eff}(\sigma_0) ~~=~~ 0\,.
\eeq
	Here we quote this as a qualitative statement and render it more precise in Section~\ref{sbps}.
	For now it is almost trivial to note that, although both superpotential functions here are multivalued, 
	there exists a {\it certain} branch on which the above difference vanishes.

\section{BPS Spectrum}
\label{sbps}
\setcounter{equation}{0}

	We now turn to the discussion of the BPS spectrum in detail, 
	with an emphasis on the strong-coupling domain.
	Surprisingly, the conclusions obtained in the strong-coupling sector will allow us
	to make implications for the weak-coupling sector as well.
	We first collect the results known and trustworthy about the CP(1) model \cite{ls1},
	as the simplest and well-studied case, and then increase $ N $.

\subsection{CP(1) Spectrum}
\label{seccp1}

	There are two kinks in the strong-coupling sector with quantum numbers
\[
 	(T,~ n)  ~~=~~ (1,~ 0) \qquad\quad \text{and}\qquad\quad (T,~ n)  ~~=~~ (1,~ 1) \,.
\]
	Here $ T $ bears the conventional meaning of the topological charge, the
	constant that multiplies $ \W_\text{eff}(\sigma_1) \,-\, \W_\text{eff}(\sigma_0) $,
	and $ n \in \mc{Z} $ is the Noether charge, connected with the angle collective
	coordinate of the unbroken U(1).

\begin{figure}
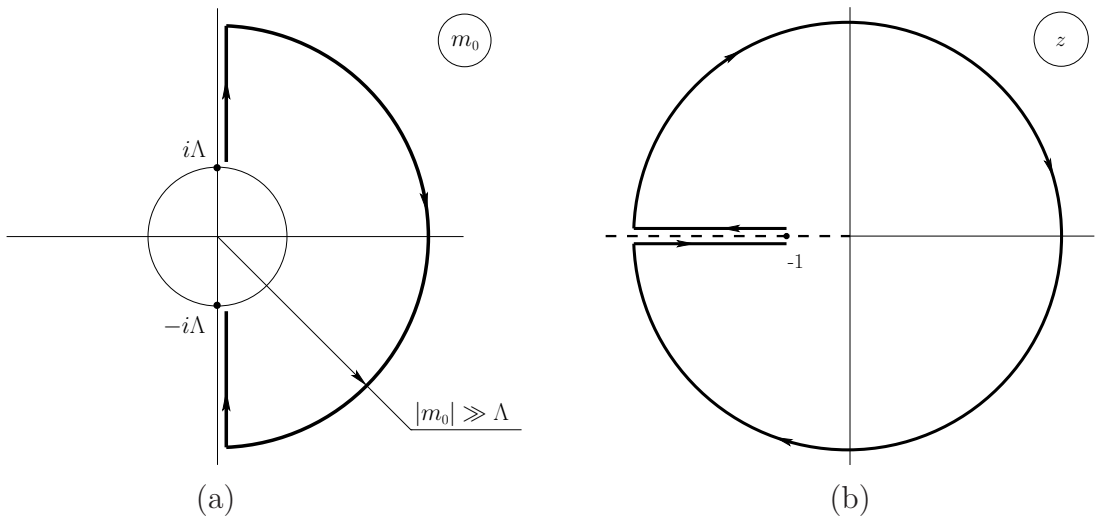

\begin{center}
\begin{tabular}{cc}
\hspace{-1.4cm}
\epsfxsize=6cm
 \epsfbox{contour_m0.epsi}
 &
\hspace{1.4cm}
\epsfxsize=5.2cm
 \epsfbox{contour_ln.epsi}
  \\
	(a)\hspace{1.7cm}  &  \hspace{2.85cm}(b)
\end{tabular}
\end{center}
\caption{(a) The large-radius contour in the $ m_0 $ plane, starting at the AD point $ i \Lambda $
	and terminating at the point $ -i \Lambda $.
	 (b) The same contour shown in the plane of 
	$ z $
	--- the argument of the logarithm in Eq.~\eqref{mcp1}.}
\label{contour_m0}
\end{figure}

	The formula for the BPS mass in the CP(1) theory is well-known. 
	We recover it from our master equations \eqref{mspec} and \eqref{unod},
\beq
\label{mcp1}
	\mbps^\text{CP(1)} ~~=~~
	\frac{1}{\pi} \lgr   2\, \sqrt{ m_0^2 \,+\, \Lambda^2\, }
			~-~ m_0\, 
			    \ln\, \frac { \sqrt{ m_0^2 \,+\, \Lambda^2\, } \,+\, m_0 }
                                        { \sqrt{ m_0^2 \,+\, \Lambda^2\, } \,-\, m_0 }
                      \rgr
	\!~~+~~
	i\,\vec{N} \cdot \vec{m}\,.
\eeq
	The 
Argyres--Douglas points here are $ m_0^\text{AD} ~=~ \pm\, i \Lambda $, and correspond to
	the vanishing square root.
	Let us trace how different states become massless at these AD points.
	Say, choose the branch of the logarithm such that at the point $ m_0 ~=~ i \Lambda $ 
	the logarithm equals $ i\pi $.
	Then the kink $ \vec{N} ~=~ (1,\, 0) $ becomes massless at this point. 
	Let us move from the point $ i \Lambda $ to $ - i \Lambda $ along a large circle, see
	Fig~\ref{contour_m0}a.
	It is easy to show that the argument of the logarithm in Eq.~\eqref{mcp1} will also follow
	a large circle, starting and terminating at $ -1 $, see Fig~\ref{contour_m0}b.
	However, this contour arrives to $ -1 $ from under the cut of the logarithm,
	and, therefore, the phase of the argument of the logarithm changes from $ i \pi $ to $ - i \pi $.
	As a result, the kink masses now shift compared to $ m_0 ~=~ +\, i \Lambda $ point,
\beq
	m_\text{BPS}^\text{CP(1)}( -\, i \Lambda ) ~~=~~ i\, m_0 ~~+~~ i\,\vec{N} \cdot \vec{m}\,.
\eeq
	One observes, that it is a different kink which becomes massless now
	(it actually is the $ (T,\, n) $ $ = (1,\, 1) $ kink, as we will see in a moment).
	It is not difficult to see that for this kink $ \vec{N} = (0,\, 1) $.

	Let us show that Eq.~\eqref{mcp1} is in one-to-one correspondence with the result from the 
	mirror theory \eqref{smirror}.
	At small $ m_0 $, the argument of the logarithm in Eq.~\eqref{mcp1} just turns into $ 1 $, 
	while the first term in the bracket tends to $ \Lambda $,
\beq
	m_\text{BPS}^\text{CP(1)} ~~\to~~ 
		\frac{1}{\pi}\, 2\Lambda  ~~+~~ i\, \vec{N} \cdot \vec{m}\,.
\eeq
	This does indeed agree with the spectrum \eqref{smirror}, if we define the two states to be
\beq
\label{scp1}
	\vec{N} ~~=~~ (1,~ 0)  \qquad\qquad\text{and}\qquad\qquad  \vec{N} ~~=~~ (0,~ 1)\,.
\eeq
	These are the two states of the strong-coupling regime. 
	How does it happen that at weak coupling one has the whole tower of states
	while at strong coupling there are only two?
	Similar to what occurs in the Seiberg--Witten theory, the states of the weak-coupling sector
	decay on the curves of marginal stability \cite{SW1,Bilal:1996sk,Bilal:1997st}.
	Only two states \eqref{scp1} survive when crossing into the strong-coupling region, 
	and those are precisely the states which become massless at the 
Argyres--Douglas points.

	The kinks \eqref{scp1} therefore must be part of the weak-coupling spectrum.
	We must be able to write a general formula for the latter, 
	keeping in mind the quasiclassical asymptotic \eqref{qclass}.
	Indeed, if one allows $ \vec{N} $ to be of the form
\beq
\label{ncp1}
	\vec{N} ~~=~~ (\, - n \,+\, 1,~ n \,)\,, \qquad\qquad\qquad n ~~\in~~ \mc{Z}\,,
\eeq
	then the two kinks \eqref{scp1} correspond to $ n \,=\, 0 $ and $ n \,=\, 1 $.
	On the other hand, at {\it large} $ n $ Eq.~(\ref{ncp1}) does reproduce the quasiclassical
	tower \eqref{qclass}.
	We stress that it is Eq.~\eqref{ncp1} that describes the exact spectrum.
	The latter was obtained based on a meaningful and single-valued formula \eqref{mspec}.
	In this illustrative example of CP(1), we stayed within one sheet of the logarithm. 

	We need to mention, that for CP(1) (and only for CP(1)) our results are not incompatible with those
	found in \cite{Dor} and described by Eq.~\eqref{qclass}, 
\beq
	(\, -n,~ n\,)\,, \qquad\qquad\qquad n ~\in~ \mc{Z}\,.
\eeq
	The reason is that the extra unity found in \eqref{ncp1} can be included into the logarithm in
	Eq.~\eqref{mcp1}.
	That will alter the sign of the expression under the logarithm, after which Eq.~\eqref{mcp1}
	will match with the analogous expression quoted in \cite{Dor} precisely.
	Already at CP(2) we will find that this coincidence is not valid.

	To summarize, the spectrum for the CP(1) theory is given by equations \eqref{mcp1} and \eqref{ncp1},
	where $ n $ runs through all integer numbers in the weak-coupling region, while it is restricted
	to $ n \,=\, 0 $ or $ n \,=\, 1 $ in the strong-coupling domain.

\subsection{General criteria}
\label{geneq}

	We can now formulate the {\it requirements} for the spectrum of BPS states in the overall
	region of the complex mass parameter $ m_0 \,$:
\begin{itemize}
\item
	{\it Quasiclassical limit} --- the spectrum at large $ m_0 $ and large excitation number $ n $
	must reproduce the semiclassical result \eqref{wkink}, \eqref{qclass}:
\beq
\label{climit}
\mbps ~~\simeq~~ \frac{N}{2\pi}\,
		( m_1 \,-\, m_0 ) \cdot \ln \frac{|m_0 |}{\Lambda}  
	    ~~+~~
	i\, n \cdot ( m_1 \,-\, m_0 )\,;
\eeq

\item
	{\it 
Argyres--Douglas point} --- the only states that survive when crossing from weak-coupling 
	into the strong-coupling region are those $ N $ states which become massless at the AD points;

\item
	{\it Mirror spectrum} --- the latter $ N $ kinks must reflect the spectrum given by mirror 
	formula \eqref{smirror} in the small $ m_0 $ limit.
\end{itemize}

	Having formulated the criteria for spectrum selection from the 
	variety of branches of the logarithms of the Veneziano--Yankielowicz superpotential,
	we can now proceed to higher $ N $.
	Of these, CP(2) will be the first nontrivial example.

\subsection{Domain of \boldmath{$ m_0 $}}

	At first we address the question of the relevant domain of variation for the parameter $ m_0 $.
	Needless to say, both Eq.~\eqref{mbpsmain} and \eqref{mspec} 
	are multivalued in the complex plane of $ m_0 $.
	The analysis of the actual complex manifold of $ m_0 $ with all its branch cuts
	is a separate problem.
	We do not need such an extended analysis for our purposes, however.
	We can limit ourselves to the domain
\beq
	-\pi ~~<~~ {\rm Arg} \left(\, m_0^N \,\right) ~~<~~ \pi\,.
\label{argu}
\eeq
	The exploitation of the criteria formulated in Sec. \ref{geneq} relies on the possibility of 
	using the formula \eqref{mspec}  
	(i) in the neighborhood of the Argyres--Douglas point;
	(ii) in the neighborhood of $ m_0 ~=~ 0 $, and 
	(iii) in the region of large $ |m_0| $ --- 
	such that there are no discontinuities ({\it i.e.} branch cuts) between these regions.

	The second goal is the calculation of the 
	curves of marginal stability, which in principle does require one to move
	over the whole complex plane of $ m_0 $.
	One can use the  physical $ 2\pi $ periodicity in the $ \theta $ angle
	to argue that the spectrum of the theory must be identical in the $ N $ sectors 
	$\left\{\,  e^{2\pi i k / N} \:~...~\:  e^{2\pi i (k+1) / N } \,\right\} $ of $ m_0 $ 
	(although there might be a monodromy between the sectors).
	The spectrum and the curves of marginal stability will therefore repeat themselves in all these
	sectors, and thus it is enough to build them in a sector $ 2\pi / N $ wide in $ \text{Arg}~m_0 $.
	
\begin{figure}
\begin{center}
\epsfxsize=8.0cm
 \epsfbox{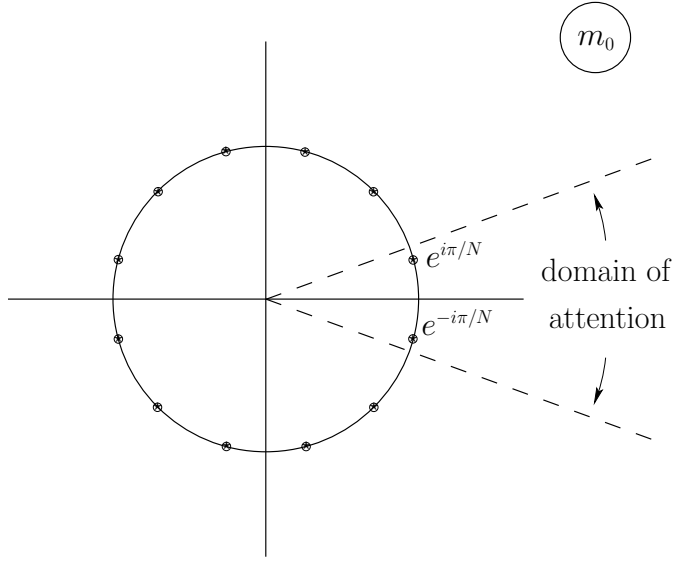}
\caption{The domain of variation of $ m_0 $ free of branch cuts}
\label{domain}
\end{center}
\end{figure}

	We choose the sector between theñå two 
Argyres--Douglas points
\beq
\label{mdomain}
	m_0^\text{AD} ~~=~~ \Lambda \, e^{i \pi / N}
	\qquad\quad
	\text{and}
	\qquad\quad
	m_0^\text{AD} ~~=~~ \Lambda \, e^{- i \pi / N}\,,
\eeq
	see Fig.~\ref{domain}.
	On the one hand the AD points are a useful reference, since the (smallest) curve of the marginal stability
	with necessity passes through the AD point (see Section~\ref{curves}).
	On the other hand, similar to what we did in CP(1), we will be able to identify the states
	that become massless at the (at least these two) AD points. 
	To this end it will be useful to be able to move from one AD point to another.
	
	Third, and perhaps most important, our function \eqref{unod} is {\it free} of the branch cuts
	in the region \eqref{mdomain}.
	This is, perhaps, the most drastic difference from the function in the right-hand side of Eq.~\eqref{mbpsmain} which 
	does have branch cuts in this region.

	One might want to  move beyond the boundaries of the domain \eqref{mdomain},
	for example, in order to approach other AD points.
	Although, as we argued, we do not need that for our purposes, we will partly be able to address the latter aspect.

\subsection{Spectrum in CP(2)}

\begin{figure}
\begin{center}
\epsfxsize=6.4cm
\epsfbox{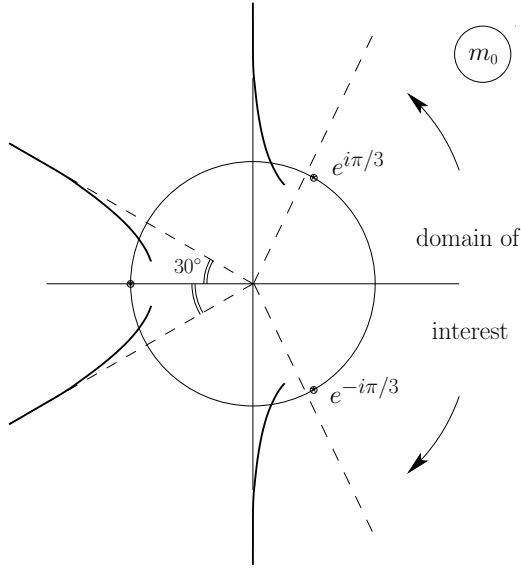}
\caption{\footnotesize
The 
Argyres--Douglas points in CP(2) theory (in the units of $\Lambda$).
Two points $ e^{\pm i \pi / N} $ lie within our area of interest.
The third one is separated from our area by branch cuts of the logarithms, shown with black solid lines.
}
\label{cp2}
\end{center}
\end{figure}

	The case of CP(2) is sufficiently straightforward so that one can analyze function $ U_0(m_0) $ in detail,
\beq
\label{unod2}
	U_0(m_0) ~~=~~ -\, \frac{1}{2\pi} \lgr e^{2\pi i / 3} \,-\, 1 \rgr 
	\bigg\{\, 3\sqrt[3]{ m_0^3 \,+\, \Lambda^3 } ~+~
		\sum_j\, m_j\, \ln \, \frac{ \sqrt[3] { m_0^3 \,+\, \Lambda^3 } \,-\, m_j } { \Lambda} \,\bigg\}\,.
\eeq
	There are three 
Argyres--Douglas points in CP(2), of which  we will be interested
	in just two, see Fig.~\ref{cp2}.
	The third one, $ m_0^\text{AD(III)} ~=~ -1 $ in the units of $ \Lambda $, 
	is separated by a pair of branch cuts of $ U_0 (m_0) $ from each of our primary AD points.
	The branch cuts start at infinity ({\it i.e.} in the semiclassical regime) and continue into the strong-coupling area,
	terminating inside the unit circle (it is to some degree a matter of convention where to terminate the branch cuts).
	Since we always prefer to move along circles of a large radius in order to stay away from the strong-coupling area, 
	these latter cuts prevent us from smoothly connecting to the third AD point.
	Nevertheless, we will still have our say about that point.

	In between the AD points $ m_0^\text{AD(I)} ~=~ e^{ i \pi / 3 } $ and 
	$ m_0^\text{AD(II)} ~=~ e^{ - i \pi / 3 } $ there are no branch cuts for \eqref{unod2},
	and one can smoothly commute between them.
	On the other hand, it can be shown that the right-hand side of Eq.~\eqref{mbpsmain} {\it does} have branch cuts in our
	area of attention, and therefore, although not impossible, when moving from one AD point to the other
	extra caution needs to be exercised.

	The function $ U_0(m_0) $ contains three logarithms, and it is easy to calculate its value
	at the AD points.
	At $ m_0^\text{AD(I)} ~=~ e^{ i \pi / 3 } $ it reduces to
\beq
	U_0(e^{ i \pi / 3 }) ~~=~~ - i\, m_0 ~~=~~ -i\, e^{ i \pi/ 3}\,. 
\eeq
	Therefore, a kink $ \vec{N} ~=~ ( 1,\, 0,\, 0 ) $ becomes massless at this point.
	On the other hand, it is straightforward to calculate the BPS mass in the small $ m_0 $ limit --- 
	the whole function \eqref{unod2} only gives a zero-order contribution in the mass, 
	the only linear term being given by $ i\, \vec{N} \cdot \vec{m} ~=~ i\, m_0 $,
\beq
\label{cp2m0}
	\mbps  ~~=~~ -\, \frac{3}{2\pi} \lgr e^{2 \pi i / 3} \,-\, 1 \rgr \Lambda ~~+~~ i\, m_0\,.
\eeq
	This is precisely one of the kinks described by the mirror formula \eqref{smirror}.

	We can now smoothly slide from the AD point $ e^{ i \pi / 3 } $ to the one $ e^{ - i \pi / 3 } $,
	along a large-radius contour ({\it i.e.} we first radially reach a circle of a large radius, and then
	sketch an arc extending clockwise to $ \text{Arg}~m_0 ~=~ - \pi / 3 $, after which, finally radially returning
	to the unit circle distance; see Fig.~\ref{cpn} where such a contour is shown for the case of general $ N $).
	One can show by tracing the corresponding contours for $ \sqrt[3]{ m_0^3 \,+\, \Lambda^3 } \,-\, m_j $ 
	that neither of the logarithms of Eq.~\eqref{unod2} steps over a branch cut. 
	At the lower AD point, one has
\beq
\label{cp2m1}
	U_0(e^{ - i \pi / 3 }) ~~=~~ -i\, m_1 ~~=~~ -i\, e^{ i \pi / 3 }\,,
\eeq
	with the same numerical value, the logarithms just got shuffled around.
	One has the kink $ \vec{N} ~=~ ( 0,\, 1,\, 0 ) $ becoming massless at this location. 
	Now similar to \eqref{cp2m0}, we obtain
\beq
	\mbps ~~=~~ -\, \frac{3}{2\pi} \lgr e^{2 \pi i / 3} \,-\, 1 \rgr \Lambda ~~+~~ i\, m_1\,.
\eeq
	This precisely matches the second of the kinks predicted by Eq.~\eqref{smirror}.

	Now how about the third kink? 
	Since we know that the right-hand side of equation \eqref{mspec} is smooth in our area of attention,
	{\it and} we know that it must account for the whole spectrum,
	we should be able to accommodate for the third kink with a certain choice of $ \vec{N} $.
	Indeed, if one allows $ \vec{N} ~=~ (0,\, 0,\, 1) $, then in the limit of small mass one has
\beq
\label{k2}
	\mbps ~~=~~ -\, \frac{3}{2\pi} \lgr e^{2 \pi i / 3} \,-\, 1 \rgr \Lambda ~~+~~ i\, m_2\,,
\eeq
	which returns us the third of the kinks of Eq.~\eqref{smirror}.
	
	Now, generally, the third AD point, $ m_0^\text{AD} ~=~ -1 $, is fenced from our area by branch cuts.
	What it in particular means, is that one cannot recover the value of $ \mbps $ in that point by smoothly sliding
	$ m_0 $ from one of our AD points, say, $ e^{ i \pi / 3 } $ over to the third one.
	It does not exclude this, however, from being performed with proper accounting for the branch cuts.
	Even though, we can still calculate the value of \eqref{unod2} in the third AD point, arguing that
	whatever the branch cuts are, they are responsible for bringing $ U_0(m_0^\text{AD(III)}) $ to the resulting form.
	We know the answer anyway, since, it is again obtained from \eqref{cp2m0} by exchanging places of the logarithms,
	and thus numerically giving the same result.
	Only now this quantity, $ e^{ i \pi / 3} $ is {\it called} differently --- $ m_2 $,
\beq
\label{cp2m2}
	U_0(-1) ~~=~~ -i\, m_2 ~~=~~ -i\, e^{ i \pi / 3}\,.
\eeq
	Since we did not follow any contour to connect this result to the Eqs.~\eqref{cp2m0} and \eqref{cp2m1},
	we further solidify Eq.~\eqref{cp2m2} with the following remark.
	In this latter equation \eqref{cp2m2}, had it been not right, we could only be off by a branch of a 
	logarithm of \eqref{unod2}.
	One can fix this branch, by approaching the third AD point from the quasiclassical area, 
	$ m_0 ~\simeq~ \sigma_0 ~=~ - \infty $, and argue that the result \eqref{cp2m2} is precisely
	what one would find.

	Independently of this, we notice that Eq.~\eqref{cp2m2} {\it indeed is} responsible for rendering the kink \eqref{k2}
	massless at $ m_0^\text{AD(III)} $.
	We therefore have ascertained the strong-coupling spectrum of CP(2), which consists of three kinks
\beq
\label{scp2}
	\vec{N} ~~=~~ 
			\quad
				\begin{array}{l}
					(\, 1,~~~   0,~~~   0 \,)\,, \\[1.5mm]
					(\, 0,~~~   1,~~~   0 \,)~~~~ \text{and} \\[1.5mm]
					(\, 0,~~ ~  0,~~~   1 \,)\,.
				\end{array} 
\eeq
	We immediately come to the following conclusion. 
	The first two states are part of the BPS tower of the weak-coupling region 
\beq
\label{ncp2}
	\vec{N} ~~=~~ (\, -n \,+\, 1,~~ n,~~ 0 \,)\,,
\eeq
	with, correspondingly, $ n ~=~ 0 $ and $ n ~=~ 1 $.
	This corresponds quasiclassically to the set of states \eqref{qclass}.
	The extra unity present in Eq.~\eqref{ncp2} can be explained 
	similarly to the one in Eq.~\eqref{ncp1}.
	Note that this unity could have equally well been placed into the second position of \eqref{ncp2},
	and yet the expression would constitute the same tower.

	There is no way, however, that the third state in Eq.~\eqref{scp2} can be found from
	Eq.~\eqref{qclass}.
	The third state is an entity that makes the CP(2) model qualitatively different from CP(1), 
	as it must be part of {\it another tower} of BPS states,
\beq
\label{another}
	\vec{N} ~~=~~ (\, -n,~~ n,~~ 1 \,)\,.
\eeq
	This sequence of states is completely new and {\it invisible} in quasiclassics!
	More exactly, quasiclassically, the two towers blend together, as the difference
	between $ n $ and $ n - 1 $ is negligible at high excitation numbers. 
	Furthermore, the contribution of the unity  at the third position 
	on the right-hand side of Eq.~\eqref{another}
	is similarly inferior to both terms in Eq.~\eqref{climit},
\beq
	\mbps ~~\simeq~~ \frac{3}{2\pi}\,
			\Delta m \cdot \ln \frac{|\Delta m|}{\Lambda}  
			    ~~+~~
			i\, n \cdot ( m_1 \,-\, m_0 )
			    ~~+~~
			i\, m_2\,,
\eeq
	and so is not seen quasiclassically either.
	The occurrence of the extra tower and the extra unity in \eqref{another} is
	a quantum phenomenon, and is the result of the fact that (any) BPS spectrum formula
	must describe the whole spectrum in any smooth region of $ m_0 $ ---
	in our case, the spectrum of states given by the mirror formula \eqref{smirror},
	and the region shown in Fig.~\ref{domain}, correspondingly.

	In summary, we found in CP(2) two towers of BPS states,
\begin{align}
\notag
	\vec{N}_{(1)} & ~~=~~ (\, -\,n_{(1)} \,+\, 1,~~ n_{(1)},~~ 0 \,)\,,\qquad\quad 
	& 
	n_{(1)} ~\in~ \mc{Z}\,,
	\\
\label{cp2towers}
	\vec{N}_{(2)} & ~~=~~ (\, ~~~~~-\,n_{(2)},~~ n_{(2)},~~ 1 \,)\,, \qquad\quad 
	& 
	n_{(2)} ~\in~ \mc{Z}\,.
\end{align}
	These towers merge with each other at large $ n $, and thus are not
	distinguishable in the quasiclassical limit.
	In the strong-coupling region, we find  three states
\beq
	\vec{N} ~~=~~ 
			~~
				\begin{array}{l}
					(\, 1,~~~   0,~~~   0 \,)\,, \\[1.5mm]
					(\, 0,~~~   1,~~~   0 \,)~~~~ \text{and} \\[1.5mm]
					(\, 0,~~~   0,~~~   1 \,)\,,
				\end{array} 
\eeq
	which form a subset of the above BPS towers: 
	$ N_{(1)} $ with $ n_{(1)} \,=\, 0 $ and  $ n_{(1)} \,=\, 1 $,
	and $ N_{(2)} $ with $ n_{(2)} \,=\, 0 $, correspondingly.

\subsection{Spectrum in CP(\boldmath{$N-1$)}}

	The crucial role which was played by equation \eqref{mspec} in the above discussion
	is a reflection of similarity of Eq.~\eqref{unod} and Eq.~\eqref{smirror}.
	Indeed, the structure of the latter two equations is identical --- 
	both formulas possess a factor of $ e^{2\pi i / N} \,-\, 1 $.
	In Eq.~\eqref{unod} this quantity multiplies a figure bracket which in the small $ m_0 $ limit
	turns into $ \Lambda $ and in  this way matches its counterpart in Eq.~\eqref{smirror}.
	The outcome of this is that, even if the branches of the general expression Eq.~\eqref{mbpsmain} were fixed by 
	some method, Eqs.~\eqref{mspec} and \eqref{unod} would still play a more prominent role than the former.
	One other confirmation of this will be given further, when we discuss  the curves
	of the marginal stability in Section~\ref{curves}.
	
	Our discussion of the CP($N-1$) theory would be in vain with  generalization of our
	previous treatment of CP(2).
	We start with two AD points within our reference area, see Fig.~\ref{domain}. 
	Choose $ m_0^\text{AD(I)} ~=~ e^{ i \pi / N } $ first.
	In order to determine which kink becomes massless at that point, one needs the value of 
	$ U_0(m_0^\text{AD(I)}) $.

	We find
\beq
\label{cpnm0}
	U_0(m_0^\text{AD(I)}) ~~=~~ -i\, m_0 ~~=~~ -i\, e^{i \pi / N}\,.
\eeq
	This equality can either be obtained by attentively looking at the logarithms of $ U_0(m_0) $ as a function, 
	without having to calculate anything, or by actual explicit summing of all the terms in $ U_0(m_0^\text{AD(I)}) $.
	We instantly find from Eq.~\eqref{cpnm0} that at $ m_0^\text{AD(I)} ~=~ e^{ i \pi / N } $ it is the kink
\beq
\label{ncpn0}
	\vec{N} ~~=~~ (\, 1,~~ 0,~~ ...,~~ 0 \,)
\eeq
	that becomes massless.
	Tending the expression \eqref{unod} to the limit of small $ m_0 $, we see that this kink has the mass
\beq
	\mbps ~~=~~ -\, \frac{1}{2\pi} \lgr e^{2\pi i / N} \,-\, 1 \rgr \Lambda ~~+~~ i\, m_0
\eeq
	near the origin.
	This obviously is one of the states of the spectrum \eqref{smirror} seen in the mirror representation.

\begin{figure}
\begin{center}
\epsfxsize=8.0cm
\epsfbox{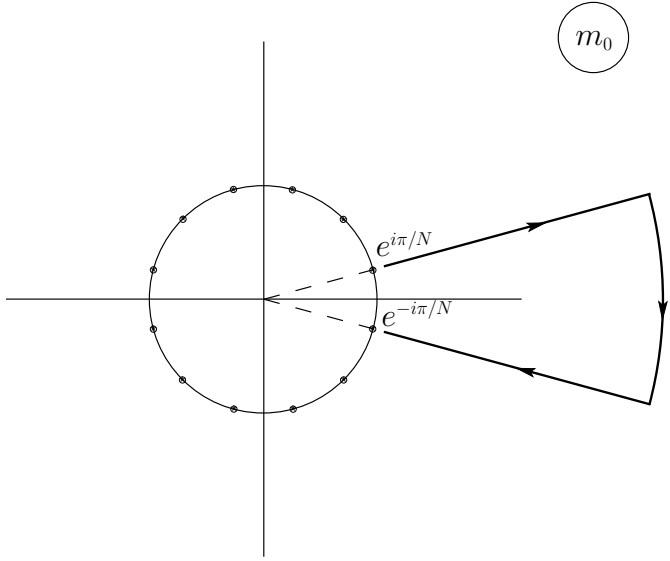}
\caption{The contour connecting two AD points $ m_0^\text{AD(I)} \,=\, e^{ i \pi / N } $ and $ m_0^\text{AD(II)} \,=\, e^{ - i \pi / N } $
in CP($N-1$) theory.}
\label{cpn}
\end{center}
\end{figure}

	The point $ m_0^\text{AD(I)} $ can be smoothly connected with $ m_0^\text{AD(II)} ~=~ e^{ - i \pi / N } $,
	see Fig.~\ref{cpn}.
	It can be shown that none of the arguments of the logarithms in the function $ U_0(m_0) $ passes through
	a branch cut, which is what we mean by smooth.
	To calculate $ U_0(m_0^\text{AD(II)}) $ one, however, can use Eq.~\eqref{cpnm0}. 
	The value of function $ U_0(m_0) $ numerically is the same in all AD points, 
	while at the point $ m_0 ~=~ e^{ - i \pi / N } $ we just endow it with a different name,
\beq
\label{cpnm1}
	U_0(m_0^\text{AD(II)}) ~~=~~ -i\, m_1 ~~=~~ -i\, e^{i \pi / N}\,.
\eeq
	This means that the kink 
\beq
\label{ncpn1}
	\vec{N} ~~=~~ (\, 0,~~ 1,~~ ...,~~ 0 \,)
\eeq
	becomes massless at $ m_0^\text{AD(II)} $.
	In the limit of small $ m_0 $, the mass of this kink takes the form
\beq
	\mbps ~~=~~ -\, \frac{1}{2\pi} \lgr e^{2\pi i / N} \,-\, 1 \rgr \Lambda ~~+~~ i\, m_1\,,
\eeq
	which, again, corresponds to one of the states described by the mirror formula \eqref{smirror}.
	The above two kinks are part of the same tower of BPS states
\beq
\label{tow1}
	\vec{N} ~~=~~ (\, -n \,+\, 1,~ n,~ ...,~ 0 \,)\,,
\eeq
	which exists in the weak-coupling region.
	Equations \eqref{ncpn0} and \eqref{ncpn1} represent the states with $ n ~=~ 0 $ and $ n ~=~ 1 $,
	respectively.

	All other AD points, although maybe disconnected from our region of interest by branch cuts, can
	still be easily dealt with, since we know the value of $ U_0(m_0) $ for any AD point,
\beq
	U_0(m_0^\text{AD}) ~~=~~ -i\, e^{i \pi / N}\,.
\eeq
	For the extra $ N - 2 $ AD points, one has, therefore,
\beq
	U_0(m_0^\text{AD}) ~~=~~ -i\, m_k\,,\qquad\qquad\quad \text{with~} k~=~ 2,\,...,\,N-1\,.
\eeq
	It does not matter which AD point produces what mass on the right-hand side, the only
	important thing is that all  other masses $ m_k $ with $ k \,\ge\, 2 $ pop up on the right-hand side.
	As it can be easily seen, in terms of the vector $ \vec{N} $,  this corresponds to the kinks 
	becoming massless at the respective AD points. The
	index $ k $ effectively shifts the unity in Eqs.~\eqref{ncpn0} and \eqref{ncpn1} further to the right.
	The full set of the states which become massless then looks as,
\beq
\label{scpn}
	\vec{N} ~~=~~ 
			\quad
				\begin{array}{l}
					(\, 1,~~~   0,~~~   0,~~~ ...,~~ 0 \,)\,, \\[1.5mm]
					(\, 0,~~~   1,~~~   0,~~~ ...,~~ 0 \,)\,, \\[1.5mm]
					(\, 0,~~~   0,~~~   1,~~~ ...,~~ 0 \,)\,, \\[0.5mm]
					\quad\quad.\quad.\quad.\quad.         \\
					(\, 0,~~~   0,~~~   0,~~~ ...,~~ 1 \,)\,.
				\end{array} 
\eeq
	Now in the limit of small masses $ m_0 $ we precisely obtain the full spectrum \eqref{smirror} 
	predicted by the mirror representation,
\beq
	\mbps ~~=~~ -\, \frac{1}{2\pi} \lgr e^{2\pi i / N} \,-\, 1 \rgr \Lambda ~~+~~ i\, m_k\,,
	\qquad\quad k~=~ 0,\,...,\,N-1\,.
\eeq

	In the weak-coupling spectrum, the states \eqref{scpn} belong to towers.
	The first two states are part of one and the same tower \eqref{tow1}, while of the rest of the states 
	each belongs to its own one.
	This agrees with a generic expectation to have $ N - 1 $ towers, according to the breaking of the 
	global SU($N$) by the masses, SU($N$) $ \to $ U(1)$^{N-1}$.
	The spectrum \eqref{cp2towers} of the CP(2) theory is then obviously extended for an arbitrary $ N $,
\begin{align}
\notag
	\vec{N}_{(1)} & ~~=~~ (\, -\,n_{(1)} \,+\, 1,~~~~\; n_{(1)},~~~~\; 0,~~~~\; 0,~~~~\; ...,~~~~\; 0 \,)\,,  
	\\[2mm]
\notag
	\vec{N}_{(2)} & ~~=~~ (\, ~~~~~-\,n_{(2)},~~~~\; n_{(2)},~~~~\; 1,~~~~\; 0,~~~~\; ...,~~~~\; 0 \,)\,,
	\\[2mm]
\label{cpntowers}
	\vec{N}_{(3)} & ~~=~~ (\, ~~~~~-\,n_{(3)},~~~~\; n_{(3)},~~~~\; 0,~~~~\; 1,~~~~\; ...,~~~~\; 0 \,)\,,
	\\
\notag
	\qquad.\qquad.
	              & \qquad.\qquad.\qquad.\qquad.\qquad.\qquad.\qquad.
	\\
\notag
	\vec{N}_{(N-1)} & ~~=~~ (\, ~~-\,n_{(N-1)},~ n_{(N-1)},~~~~\, 0,~~~~\; 0,~~~~\; ...,~~~~\; 1 \,)\,,
\end{align}
	with all $ n_{(k)} $ integer numbers.
	Obviously, in the quasiclassical limit, these towers reproduce the asymptotics \eqref{qclass}, and 
	therefore satisfy all three criteria which we listed in the end of Section~\ref{seccp1}. 
Note, that the total number of states in the physical weak-coupling spectrum is $(N-1)\,\mc{Z}$, much less then
the total multiplicity $\mc{Z}^{N}$ allowed by \eqref{mspec}.

	Equations \eqref{cpntowers} and \eqref{scpn} are our main results for the spectrum of the CP($N-1$) 
	model
	in the weak- and strong-coupling regimes, respectively.
	They exhibit a drastic difference with what was thought of the spectrum of CP($N-1$) earlier \cite{Dor},
	when it was argued that only one tower of the BPS states existed (for a given pair of vacua, of course). 
	We stress that the emergence of the extra $ N - 2 $ towers is a pure quantum effect which could not
	be anticipated in the quasiclassical theory.
	All of $ N - 1 $ towers of states blend and become degenerate in the quasiclassical limit, making it
	hard to resolve them apart.

\section{Curves of Marginal Stability}
\label{curves}
\setcounter{equation}{0}

	The fact that there are $ N - 1 $ towers of BPS states means that there should be 
	$ N - 1 $ curves on which those towers collapse. 
	Looking at Eq.~\eqref{cpntowers} one can tell that the first of them is special,
	just by its appearance ---
	the unity stands in one row with $ n_{(1)} $ (it is a mere matter of convention
	whether to write this unity in the first or in the second position).
	We will find that this tower is also special for an objective reason --- namely,
	its decay curve will necessarily pass through the AD point, while those of the other towers will not.
	For the same token, it will also be the {\it innermost} curve.
	That is, inside this decay curve, only strong-coupling states \eqref{scpn} exist.

	We provide an important technical remark on the graphical illustrations in the following discussion.
	We will choose to draw curves of marginal stability in the plane of $ m_0^N $ rather
	than in that of $ m_0 $.
	Because of $ 2\pi $-periodicity in $ \theta $-angle, a curve sketched in the $ m_0 $ plane 
	repeats itself $ N $ times in each of $ 2 \pi / N $ sectors of the argument of $ m_0 $.
	This way, drawing a curve in the $ m_0^N $ plane is as informative. 
	We call for attention however, that when drawing multiple curves, we will have 
	to do special rescaling in the $ m_0^N $ plane, for the sake of fitting multiple figures in one plot.

	Having the spectrum of the theory at hand, it does not cost an effort to write the equations 
	for the curves of marginal stability.
	Each such equation needs to rephrase the condition that one of the towers \eqref{cpntowers}
	completely decays.
	Let us write explicitly the expression for the mass of a BPS state as per the spectrum \eqref{cpntowers},
\beq
\label{mtower}
	\mbps ~~=~~ U_0 (m_0) ~~+~~ i\, n_{(k)} \cdot ( m_1 \,-\, m_0 ) ~~+~~ i\, m_k\,,
	\qquad k ~=~ 1,\,...,\, N-1\,.
\eeq
	Here for sake of convenience we redefined the U(1) charge $ n_{(1)} $ 
	in \eqref{cpntowers} with $$ n_{(1)} \,\to\, n_{(1)} \,+\, 1 \,.$$
	In terms of the expression for the mass, the ``tower''  
	is given by the term $$ i\, n_{(k)} \cdot ( m_1 \,-\, m_0 ) \,.$$
	For the tower to decay, the remainder in the right-hand side of Eq.~\eqref{mtower} must be
	in phase with the latter term, or
\vspace{2mm}
\beq
\label{cms}
	\text{Re}~~ \frac{ U_0 (m_0) ~~+~~ i\, m_k }
                         { m_1 ~~-~~ m_0 }    ~~=~~ 0\,. \\[1.5mm]
\eeq
	This is the CMS equation.
	We obtain $ N - 1 $ curves here by letting $ k $ run from 1 to $ N - 1 $.

	Now, we can make a few assertions on CMS before starting drawing them:
\begin{itemize}
\item
	The curves \eqref{cms} either do not intersect or they completely overlap;

\item
	The {\it primary} curve corresponding to $ k \,=\, 1 $ (and  only this curve) passes through the AD point.
\end{itemize}
	The first assertion is seen from Eq.~\eqref{cms} directly.
	If the curves with $ k \,=\, p $ and $ k \,=\, q $ happen to intersect somewhere, then
\beq
\label{overlap}
	\frac{ \,m_p ~~-~~ m_q\, }
             { \,m_1 ~~-~~ m_0\, } ~~\in~~ \mc{R}
\eeq
	at that place.
	But this ratio does not depend on the absolute value of $ m_0 $, nor on its phase, so
	it will remain real along both of the curves, which for this reason will have to completely match.
	We will find that this does happen all along.

	The second assertion, although obvious as well, deserves more attention.
	To see that the curve $ k \,=\, 1 $ passes through the AD point, we rewrite Eq. (\ref{cms}) as
\vspace{2mm}
\beq
\label{cms0}
	\text{Re}~~ \frac{ U_0 (m_0) ~~+~~ i\, m_0 }
                         { m_1 ~~-~~ m_0 }    ~~=~~ 0\,, \\[1.5mm]
\eeq
	where the substitution $ m_1 \,\to\, m_0 $ in the numerator obviously does not change the condition
	(in fact, Eq.~\eqref{cms0} is exactly what we would have obtained from the spectrum \eqref{cpntowers}
	if we had not done the shift of $ n_{(1)} $ above).
	But we calculated the value of $ U_0(m_0) $ at the AD point $ m_0^\text{AD(I)} $ in Eq.~\eqref{cpnm0},
	which shows that the CMS condition is trivially met there.
	Then the curve has to pass through all AD points.

\begin{figure}
\begin{center}
\epsfxsize=6.0cm
\epsfbox{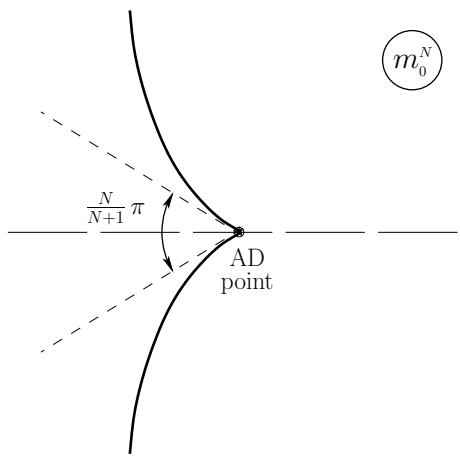}
\caption{The cusp of the primary decay curve at the AD point in the $ m_0^N $ plane.}
\label{cusp}
\end{center}
\end{figure}

	In the $ m_0^N $ plane there is just one such point, and, since $ \sigma_p $ vanishes in it, 
	we can expand the above condition in $ \sigma_0/m_0 $ in its neighborhood.
	We have,
\beq
\label{alpha}
	\text{Re}~~ \sum_{r \,>\, 0}\: \frac{ \alpha^{r N \,+\, 1} }
                                          {\:  rN ~+~ 1 \:}          ~~=~~ 0\,,
	\qquad\qquad \alpha ~=~ \frac{\sigma_0}{m_0}\,.
\eeq
	This function, up to a constant, is actually known as the so-called Hurwitz--Lerch transcendent, 
	and is a special case of the hypergeometric function. 
	For our purposes, however, we only need the leading order term of it,
\beq
	\qquad\qquad\qquad\qquad
	\text{Re}~~ \alpha^{ N \,+\, 1 } ~~=~~ 0\,, \qquad\qquad \text{for~~} \alpha ~\ll~ 1\,.
\eeq
	Solving this equation gives us the {\it angle} at which the $ k \,=\, 1 $ curve passes through the AD point,
\beq
\label{angles}
	\qquad\qquad\quad        
	\phi ~~=~~ \left \{ ~
		\begin{array}{l}
		   { \displaystyle
		    +\: \frac{ N \,+\, 2 }
                             { N \,+\, 1 }\cdot \frac{\pi}{2}\,,
                   }
		    \\[4mm]
		   { \displaystyle
		    ~~~~~~~~~0\,, \qquad~~ \text{(for CP(1) only)}
                   }
		    \\[3mm]
		   { \displaystyle
		    -\: \frac{ N \,+\, 2 }
                             { N \,+\, 1 }\cdot \frac{\pi}{2}\,,
                   }
		\end{array}
		\right.
\eeq
	where, as customary, we measure the angle from the real positive direction counter-clockwise.
	Equation \eqref{angles} tells us that for $ N \,>\, 2 $ there are two opposite angles, and 
	the decay curve has a {\it cusp} at the 
Argyres--Douglas point, see Fig.~\ref{cusp}.
	As $ N $ becomes larger, the opening angle of the cusp increases, and ultimately,
	when $ N $ is taken infinitely large the curve becomes smooth.

	CP(1) theory is special, as its curve has three angles $ +120^\circ $, $ -120^\circ $ and $ 0^\circ $ instead of two.
	The third angle corresponds to the extra flat part $ [-1,\, 0] $ of the curve sticking out of the AD point,
	see Fig.~\ref{ccp1}.
	Otherwise, the CP(1) case is the simplest, and so we start the illustrative part of our discussion with this theory.

\subsection{The decay curve in CP(1)}

	There is just one curve of marginal stability in CP(1) and it is the primary one.
	It represents a sharp boundary between the areas of weak and strong coupling spectra. 
	This curve and its features have been discussed in the literature \cite{ls1,Olmez}, so we
	just merely reproduce it with Eq.~\eqref{cms}.
\begin{figure}
\begin{center}
\epsfxsize=7.5cm
\epsfbox{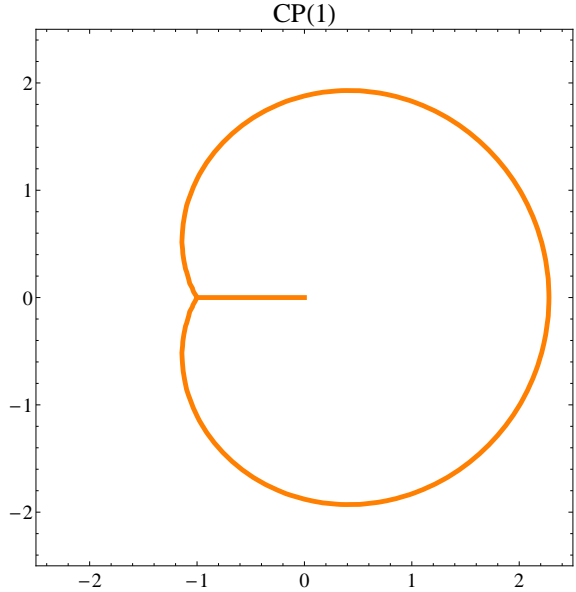}
\caption{The curve of marginal stability in CP(1) theory ($ m_0^2 $ plane), Ref. \cite{ls1}.}
\label{ccp1}
\end{center}
\end{figure}
	Figure~\ref{ccp1} shows the curve  in the plane of $ m_0^2 $. 
	The whole graph is presented in units of $ \Lambda^2 $.
	We restate the known facts that the curve has a cusp at the AD point, where also
	an extra part $ [-1,\, 0] $ of the curve connects.
	All three lines meet at the AD point at an angle $ 120^\circ $ with respect to each other.
	The real interval $ [-1\,, 0] $ is the analytical solution to the CMS condition.
	There is nothing like that for any other CP($N-1$) theory, all other curves are
	just curves.

\subsection{CMS in CP(2) theory}

\begin{figure}
\begin{center}
\epsfxsize=7.5cm
\epsfbox{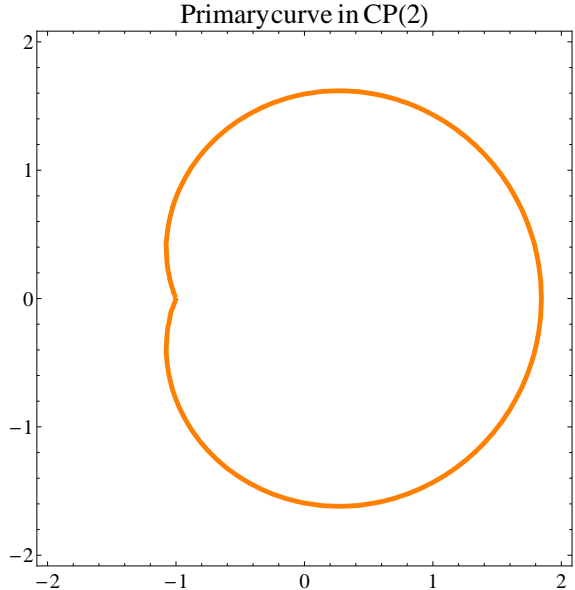}
\caption{The primary decay curve in the CP(2) theory ($ m_0^3 $ plane).} 
\label{pccp2}
\end{center}
\end{figure}

	The CP(2) theory features two curves --- one primary and one secondary. 
	The primary curve, as we established, has a cusp at the AD point with the
	opening angle $ 135^\circ $, see Fig.~\ref{pccp2}.

\begin{figure}
\begin{center}
\epsfxsize=7.5cm
\epsfbox{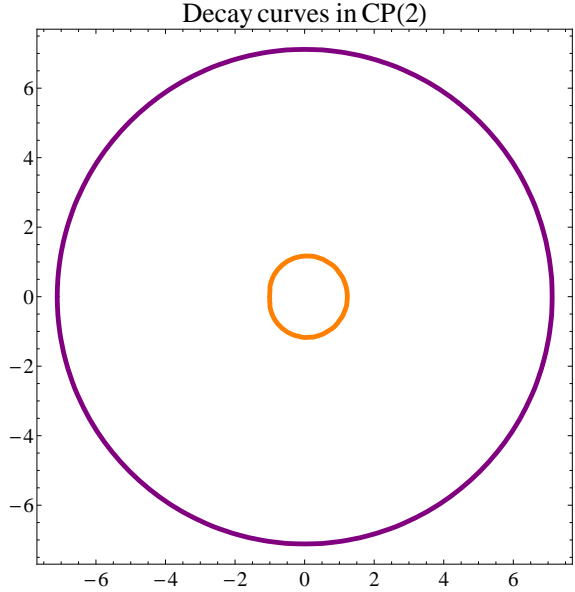}
\caption{Both decay curves in the CP(2) theory (compressed $ m_c $ plane, see text).} 
\label{ccp2}
\end{center}
\end{figure}

	The second curve is a circle-like loop with a radius of approximately $ 361\, \Lambda^3 $ 
	(as we will see, it {\it is} a circle to a very good accuracy).
	This is a very large circle, compared to the primary curve which is of the size $ \Lambda^3 $.
	In order to plot them both on the same graph, we rescale the {\it radial} direction of 
	$ m_0^N $ by taking the $ N $-th root from its absolute value, while leaving the 
	phase of $ m_0^N $ as  is, 
\beq
\label{mc}
	\big| m_c \big| ~~\equiv~~ \big| m_0^N \big|^{1/N} \,, \qquad\qquad \text{Arg}\:m_c ~~=~~ \text{Arg}\: m_0^N\,.
\eeq
The subscript $c$ stands for compressed.
	Figure~\ref{ccp2} shows the two curves in the rescaled $ m_0^3 $ plane.
	This compression will appear useful for the large $ N $ case.
	Such a transformation, certainly, distorts the cusp making it less expressed.
	But, Figure~\ref{pccp2} assures us that it is there, and Figure~\ref{cusp} tells us that
	it is there for all CP($N-1$).

	The radius of the external curve, $ 361\, \Lambda^3 $ is found to be rather large.
	However, in the compressed $ m_c $ plane (Fig.~\ref{ccp2}), and, equivalently, in the $ m_0 $ plane, 
	the curve has the size approximately $ 7.12\, \Lambda $. 
	We will find later, that the maximal radius of all curves (in the large $ N $ limit)
	is $ e^2 \, \Lambda $, which amounts to $ 7.39\, \Lambda $.
	Thus, already the secondary curve of CP(2) nearly saturates the maximum size.

\subsection{Larger-\boldmath{$ N $} theories}

\begin{figure}
\begin{center}
\epsfxsize=7.5cm
\epsfbox{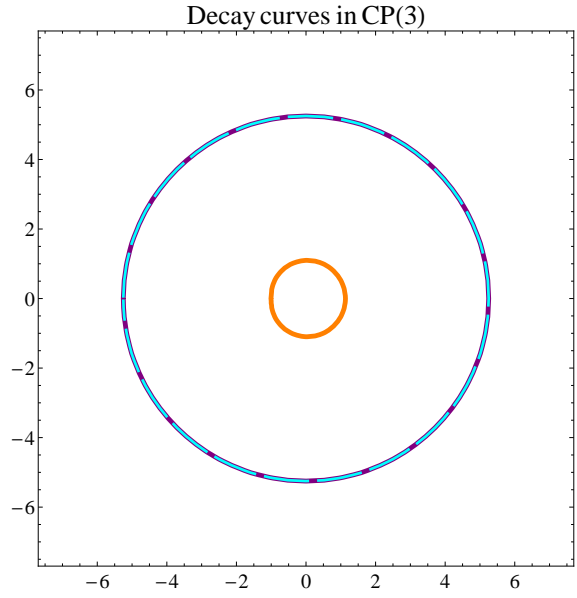}
\caption{\small Three decay curves in the CP(3) theory (compressed $ m_c $ plane, see Eq.~\eqref{mc}).} 
\label{ccp3}
\end{center}
\end{figure}

	For the CP(3) theory, the curves are shown in Fig.~\ref{ccp3}.
	Again, we plot the curves in the plane where $ m_0^N $ was radially compressed to $ m_c $.
	The two outer curves in CP(3) overlap, as a consequence of Eq.~\eqref{overlap},
\beq
	m_2 ~~-~~ m_3  ~~=~~ m_1 ~~-~~ m_0\,.
\eeq
	The cusp in the primary curve is still present, although is flattened in Fig.~\ref{ccp3} due to compression.

\begin{figure}
\begin{center}
\epsfxsize=7.5cm
\epsfbox{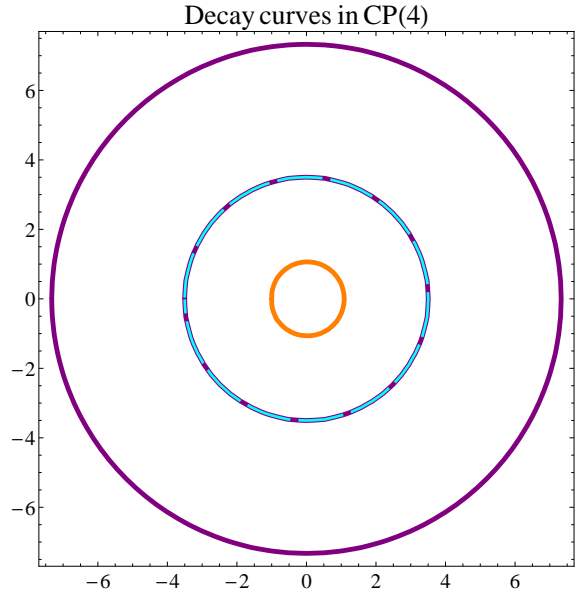}
\caption{\small The decay curves in the CP(4) theory (compressed $ m_c $ plane, see Eq.~\eqref{mc}).
External radius is $ 7.33\, \Lambda $.} 
\label{ccp4}
\end{center}
\end{figure}

\begin{figure}
\begin{center}
\epsfxsize=7.5cm
\epsfbox{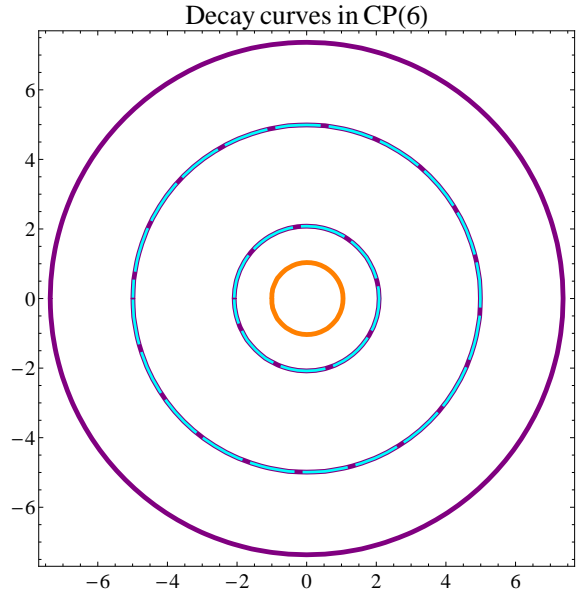}
\caption{\small The decay curves in the CP(6) theory (compressed $ m_c $ plane). Outer radius is $ 7.37\, \Lambda $.} 
\label{ccp6}
\end{center}
\end{figure}

\begin{figure}
\begin{center}
\epsfxsize=7.5cm
\epsfbox{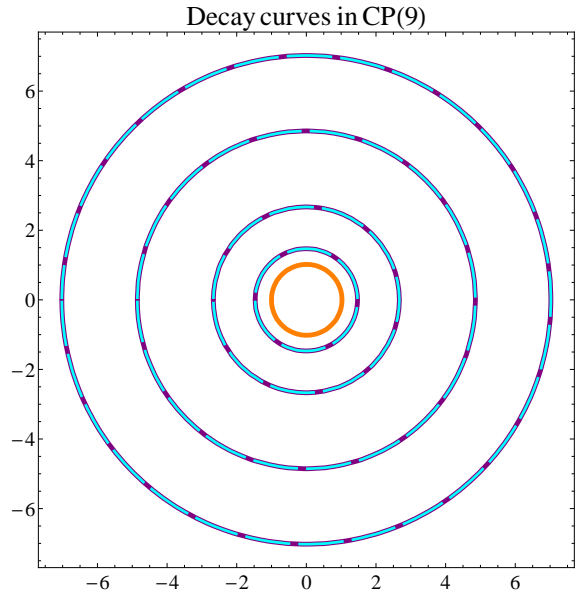}
\caption{\small The curves in CP(9) theory (compressed $ m_c $ plane). Outer radius is $ 7.02\, \Lambda $.}
\label{ccp9}
\end{center}
\end{figure}

	For CP(4) the curves are shown in Fig.~\ref{ccp4}. 
	The overlapping curves are shown with portioned lines. 
	The radius of the outer most curve is $ 7.33 \Lambda $, which is very close to the upper limit!
	Figures~\ref{ccp6} and \ref{ccp9}, as an illustration, show the curves for the CP(6) and CP(9).

	Let us briefly review the main features of these plots. 
	We observe that all decay curves look as nearly perfect circles.
	We do not set the goal here to rigorously prove that they are, 
	but we are able to show that the secondary curves for a few first theories are circles to a 
	good accuracy and have no reason to believe that this is not so for larger $ N $.
	More detailed analysis of their shape lies outside the scope of this paper.

	We know that the primary curves are not circular because of the cusp, and if in our figures 
	they look round, this is just an illusion caused by the compression. 
	For example, for CP(9) with $ N $ as large as $ 10 $, the cusp opening angle would be $ 164^\circ $, 
	which although close to $ 180^\circ $, would still be quite noticeable.
	It is true, however, that at larger $ N $ even the primary curves turn into circles (that is,
	in any plane).
	We will discuss this in the next subsection.

	However, the secondary curves are circular even for small $ N $.
	Using the $ (\Lambda/m_0)^N $ expansion of the CMS condition \eqref{cms} it is possible to prove
	the following statement.
	If it is known, that a particular curve passes at a large distance from the origin at least
	at one point, then such a curve must be a circle as perfect as $ (\Lambda / m_0)^N $.
	For example, in CP(2), Fig.~\ref{ccp2}, the outer curve is a circle with the accuracy about $ 1/360 $.
	This remark allows us to see that the secondary curves are round for a few starting CP($ N-1 $) theories,
	just by looking at the location where the curves cross the real axis.
	For the primary curves this statement obviously does not apply since they pass through the AD point 
	at a unit distance from the origin.

	For larger $ N $, the $ k ~=~ 2 $ curves come closer and closer to the primary one, 
	and one might suspect that they start losing their shape.
	For example, the radius of the $ k \,=\, 2 $ curve in CP(14) theory is only $ \sim~14 $ in units of $ \Lambda^{15} $.
	A deviation from a circular figure at the level of $ 1/14 $ would be quite noticeable.
	However, a different effect takes over, which helps the decay curves stay ``fit''.

\subsection{Curves in the large-\boldmath{$ N $} limit}

	We now consider the question of the form of the decay curves in the limit of large $ N $.
	It turns out that the analysis greatly simplifies.
	All the curves turn into circles of a certain radius. 

	As usual, the primary curves are a separate topic. 
	Consider the curve in the neighborhood of the AD point, where it is described by Eq.~\eqref{alpha},
\beq
\label{alphaagain}
	\text{Re}~~ \sum_{r \,>\, 0}\: \frac{ \alpha^{r N \,+\, 1} }
                                          {\:  rN ~+~ 1 \:}          ~~=~~ 0\,,
\eeq
	and discard the unity in the denominator.
	The sum then reassembles into a logarithm,
\beq
	\text{Re}~\, \frac{\sigma_0}{m_0}\,\log \Big(\, 1 \:-\: \frac{\sigma_0^N}{m_0^N} \,\Big)  ~~=~~ 0\,,
\eeq
	where we have replaced $ \alpha $ with its definition.
	As $ N $ goes to infinity, the vacuum $ \sigma_0 $ approaches the value $ m_0 $, and we can replace the ratio
	$ \sigma_0 / m_0 $ in front of the logarithm with unity.
	The logarithm itself can be transformed into
\beq
	\text{Re}~\, \log\, \Big(\, -  \frac{\Lambda^N}{m_0^N} \,\Big) ~~=~~ 0\,.
\eeq
	This equation is trivially solved if $ (m_0/\Lambda)^N $ is a pure phase.
	The curve then has to be a circle everywhere, not only in the region of validity of expansion \eqref{alphaagain}.
	We thus have shown that at large $ N $ all primary curves tend to a circle of unit radius (in units of $ \Lambda^N $).

	A similar idea, but in a different realization is used to reveal the shapes of the secondary curves.
	In the region of space where $ | \sigma_0 | \,>\, | m_0 | $ (in particular, at large positive $ m_0 $) one can expand 
	the CMS condition in $ 1/\alpha $.
	This gives,
\beq
	-~\: \text{Re}~~
	\sum_{r \,\ge\, 0}\: \frac{   \alpha^{ -( r N \,-\, 1) }   } 
                                  {\:         r N \,-\, 1        \:} ~=~
	\frac{ 2 \pi } { N }\,
	\frac{   \cos\, \frac{ 2\, k \,-\, 1 } { N }\, \pi   }
             {   2\, \sin\, \frac{ \pi }{ N }   }  
	\,,
\eeq
	where $ k $, again, is the number of the curve. 
	At large $ N $ we can again drop the unity in the denominator on the left-hand side, 
	which turns the sum into a logarithm.
	Also we expand the sine in the right-hand side to the leading order in $ 1/N $.
	We have
\beq
	1 ~~-~~ \ln\, \big| \sigma_0 \,/\, \Lambda \big|  ~~=~~ \cos\, \frac{\scriptstyle 2\, k \,-\, 1 } {\scriptstyle N }\, \pi \,.
\eeq
	Replacing $ \sigma_0 $ with $ m_0 $ with an exponential accuracy, we arrive at the formula,
\beq
\label{cos}
	\big| m_0 \big| ~~=~~ e^{ 1 \;-\; \cos\, \frac{\scriptstyle 2\, k \,-\, 1 } {\scriptstyle N }\, {\scriptstyle \pi} }\,,
	\qquad\qquad 
	k ~=~ 1,\, ...,\, N\,-\,1\,,
\eeq
	in units of $ \Lambda $.
	Even though this formula has been derived in the assumption of large $ N $, it {\it qualitatively} gives 
	a reasonable answer even for $ N $ as low as three!
	
	The qualitative features, which are obeyed in all CP($ N - 1 $) theories at large $N$ are as follows.
	The curves come in overlapping pairs, as given by the cosine in Eq.~\eqref{cos}.
	The {\it minimum} radius of the CMS is one, and is saturated by the primary curve --- this fact we already know.
	The {\it smallest secondary} curves correspond to $ k \,=\, 2 $ and $ k \,=\, N - 1 $.
	Their size depends on $ N $.
	Finally, the {\it maximum} size is 
\beq
	m_0^\text{max} ~~=~~ e^2\,,
\eeq
	measured in units of $ \Lambda $, and is reached by the 
	$ k \,=\, \frac{ N \,+\, 1 } { 2 } $ or the
	$ k \,=\, \frac{ N \,+\, 1 \,\pm\, 1 } { 2 } $
	curves, depending on parity of $ N $.
	For odd $ N $, the largest curve does not have an overlapping party.

	Interestingly enough, although these qualitative results were inferred from the large-$ N $ formula \eqref{cos},
	they are still valid for small $ N $ theories!
	In particular, $ e^2 $ seems to be the absolute limit for the size of all of the curves 
	(the deviation is that this limit is not attained at small $ N $, but some curves do come very close 
	as we saw above).
	Also, the pairing of overlapping curves described by the cosine appears to be correct for any $ N $.

\begin{figure}
\begin{center}
\epsfxsize=7.2cm
\epsfbox{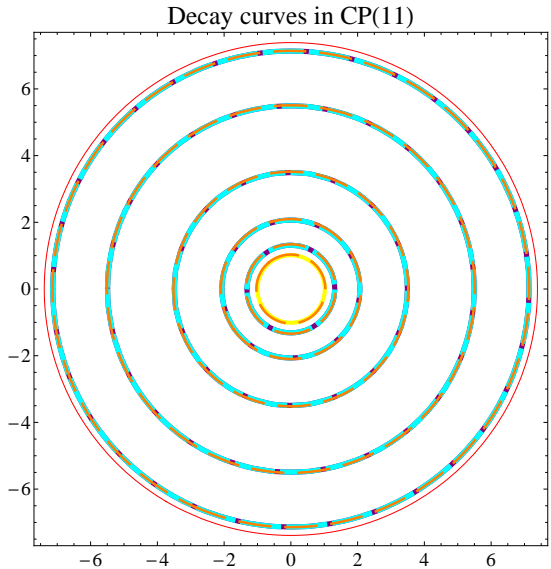}
\caption{\small The curves in the CP(11) theory (compressed $ m_c $ plane).}
\label{ccp11}
\end{center}
\end{figure}

	We illustrate the large $ N $ limit formula \eqref{cos} in Fig.~\ref{ccp11}.
	The external thin circle envelopes the overall $ |m_0| \,=\, e^2 $ size of the figure.
	The circles of radii determined by Eq.~\eqref{cos} are shown with thin dashed lines, which perfectly overlay
	the numerical curves. 
	In fact, the latter formula has a good agreement with CMS curves already for $ N \,=\, 8 $, while, as we mentioned,
	in overall it shows the right tendency already for $ N $ as low as three.

\begin{figure}
\begin{center}
\epsfxsize=7.5cm
\epsfbox{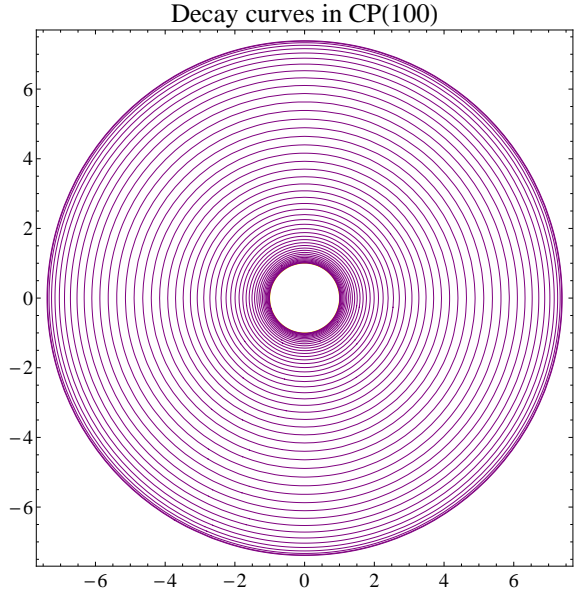}
\caption{\small The curves in the CP(100) theory (compressed $ m_c $ plane).}
\label{ccp100}
\end{center}
\end{figure}

	As we increase $ N $, the decay curves fill in the whole interval 
\beq
	\big| m_0 \big| ~~\in~~ 1~~ ...~~e^2\,.
\eeq
	Formula \eqref{cos} predicts how the curves lay into this interval.
	As a concluding illustration, Fig.~\ref{ccp100} shows the curves of the CP(100) theory.

\section{Prospects for the Spectrum in Quasi-Classics}
\label{quasi}
\setcounter{equation}{0}

	We have used strong-coupling techniques, such as the mirror symmetry, in order to
	establish the form of the BPS spectrum,
\beq
\label{prosp}
	\mbps ~~=~~ U_0 (m_0) ~~+~~ i\, n_{(k)} \cdot ( m_1 \,-\, m_0 ) ~~+~~ i\, m_k\,,
	\qquad k ~=~ 1,\,...,\, N-1\,.
\eeq
	At weak coupling, it must also be possible to see the features of this result, 
	via semiclassical methods.
	At large $ m $, function $ U_0(m_0) $ gives the canonical logarithmic contribution as shown in Eq.~\eqref{climit},
	while the mass difference $ m_1 \,-\, m_0 $ constructs the towers.
	These two contributions are expected.
	A very non-trivial prediction of Eq.~\eqref{prosp} is the last term, which distinguishes the
	different towers.
	If one re-writes all relevant equations in terms of the mass differences, then
\beq
	m_k ~~=~~ \frac{1}{N}\, \sum_j\, (\, m_k ~~-~~ m_j \,)\,,
\eeq
	since the average mass is zero in our case. 
	This shows in fact that the last term in Eq.~\eqref{prosp} represents a fractional U(1) charge.
	
	The phenomenon of the occurrence of a fractional charge is well-known \cite{Jackiw:1975fn,ls1,Shifman:2005rs,Shifman:2007ce}, 
	and is due to the presence of fermions.
	In particular, for CP(1) one has a half-unit contribution to the central charge \cite{ls1}.
	We observe this one half by re-writing the last term of Eq.~\eqref{prosp} in terms of the
	mass difference
\beq
	i\, m_1 ~~=~~ - i\, m_0 ~~=~~ \frac{1}{2}\, i\, \Delta m\,,
\eeq
	which acts as a half-integer addition to $ n_{(1)} $.
	Therefore in the CP(1) model there is an agreement between our spectrum and the known quasiclassical results.
	Recovery of the weak-coupling spectrum in a general CP($N-1$) theory via semiclassical methods 
	is an important and non-trivial problem deserving a separate study. 
	We plan to address this topic elsewhere.

\section{Conclusion}
\label{conclu}
\setcounter{equation}{0}

	We demonstrated that the weak-coupling spectrum of the CP($N-1$) theory with 
	twisted $ \mc{Z}_N $ masses is considerably richer than was thought before.
	In particular, there are $ N - 1 $ infinite towers of the BPS states. 
	Our analysis relied on three important facts known about CP($N-1$): the
	strong-coupling spectrum only includes $ N $ states; these states become massless at 
	the Argyres--Douglas points; and the quasiclassical spectrum contains a tower of states 
        with integer $ U(1) $ charges. The knowledge of the exact superpotential is not 
        sufficient as it is  ambiguous. However, if one fixes this ambiguity at least in some 
        region of the parameter space, then the superpotential must still describe the states 
        in the strong-coupling domain.
	It is possible to trace these states all the way from the weak-coupling region 
	through the AD points and into the strong-coupling domain, where their masses can be fixed
	by comparing with the mirror representation.
	We emphasize that we observe $ N - 1 $ towers instead of just one, which is naturally
	explained by the fact that the global SU($ N $) symmetry is broken down to $ N-1 $ copies of U(1),
	and the central charges of the model include terms proportional to $N-1$ Noether charges.
	However, only one of these towers is seen quasiclassically. 
	Furthermore, these $ N - 1 $ towers blend together in the quasiclassical limit, 
	making it hard to anticipate their existence from the semiclassical analysis alone.
		
	Having obtained the spectrum, it is easy to construct the curves of the marginal stability.
	We find $ N - 1 $ such curves, one per each BPS tower.
	One of the curves, which we refer to as primary, is special as it  
	passes through the AD point, with necessity. 
	Inside this curve, only  $ N $ stable states of the strong-coupling spectrum survive.
	We also  considered the large-$ N $ limit and argued that all curves tend to a
	circular shape, with the radius lying in between $ 1 $ and $ e^2 $, in units of $ \Lambda $.

	We note that we   only analyzed the decay curves of elementary states. 
	In principle, nonelementary kinks  have their own series of curves \cite{LeeYi}.
	Also, the theory must have the fermion-soliton bound states \cite{Dorey:1999zk}, 
	for which there will exist CMS as well.

As was shown in \cite{Dor},
the BPS spectrum of dyons (at the singular point on the Coulomb branch 
in which  $N$ quarks become massless) of \ntwo  four-dimensional supersymmetric QCD
  with the U$(N)$ gauge group and $N_f=N$ quark flavors, identically
coincides with the BPS spectrum in the two-dimensional  CP$(N-1)$ model.
 The reason for this coincidence was revealed 
in \cite{SYmon,HT2}. 
	The confined 't Hooft--Polyakov monopoles of the four-dimensional theory are represented by junctions of two different non-Abelian strings. 
	The effective theory on the world sheet of the non-Abelian string is the two-dimensional CP$(N-1)$ model \cite{HT1,ABEKY,SYmon,HT2}. 
	Confined monopoles of the bulk theory are seen in the world-sheet theory as kinks of the CP$(N-1)$ model. 
	This ensures the  coincidence of the BPS spectra in both theories, see \cite{SYmon,HT2} for more details.

The above coincidence implies that we can use the  results for BPS spectrum of the CP$(N-1)$ model obtained in this paper to construct the physical spectrum of the BPS dyons in \ntwo supersymmetric QCD in the particular vacuum 
in which  $N$ quarks become massless.
More specifically, the masses of the BPS states in four dimensions are given by the periods of the 
	Seiberg--Witten differential, and have exactly the same form as the 
	central charge of the sigma model \eqref{mbpsgen} written in terms of the 
Veneziano--Yankielowicz superpotential.
	The masses of dyons are determined by the contours that encircle the branch points of the
	Seiberg--Witten curve.
	The spectrum that we derived in the sigma model gives the prescription on how to build 
	contours that correspond to stable BPS states. In particular, our result \eqref{cpntowers}
for the weak-coupling spectrum of the CP$(N-1)$ model suggests that the dyon electric charges in the bulk theory,
besides the towers built on the roots of the gauge group also contain contributions determined by its weights,
cf.~\cite{FrHoll}.
	The question of the correspondence between the spectra of the BPS states and 
	the decay curves of dyons of the two theories calls for a special investigation.
	
	\section*{Acknowledgments}
	\addcontentsline{toc}{section}{Acknowledgments}
 The work of MS was supported in part by DOE
Grant No. DE-FG02-94ER408. 
The work of PAB is supported by the DOE Grant No. DE-FG02-94ER40823.
The work of AY was  supported
by  FTPI, University of Minnesota,
by RFBR Grant No. 09-02-00457a
and by Russian State Grant for
Scientific Schools RSGSS-65751.2010.2.

\end{document}